\documentclass[a4paper,11pt]{article}
\usepackage{jheppub} % for details on the use of the package, please see the JINST-author-manual
\usepackage{lineno}
\usepackage[normalem]{ulem}

\title{\boldmath A fluid dual to the charged large \texorpdfstring{$D$}{D} membrane paradigm }

\author[a]{Supratim Halder,}
\author[a]{Manu Kurian,}
\author[a]{and Mangesh Mandlik}

\affiliation[a]{Department of Physics, Indian Institute of Technology
(Indian School of Mines) Dhanbad,\\ Jharkhand 826004, India}

\emailAdd{23dr0184@iitism.ac.in} 
\emailAdd{manukurian@iitism.ac.in}
\emailAdd{mandlik@iitism.ac.in}

\abstract{According to the formulation of the charged large $D$ membrane paradigm, an arbitrary dynamic black hole solution to a theory of gravity with a $U(1)$ gauge field is dual to the dynamics of a membrane in a non-gravitational background. This membrane is endowed with a stress-energy tensor and a charge current, whose conservation equations govern its dynamics. In this work, we demonstrate that the dynamics of these membrane configurations (at the leading nontrivial order in $1/D$) can be mapped to a relativistic charged fluid, establishing a correspondence for asymptotically flat black holes with a particular class of fluid systems. Unlike the standard AdS/Hydrodynamics correspondence, this dual fluid does not reside on an asymptotic boundary, but is localised strictly on the non-gravitational membrane worldvolume. By evaluating the system in both the Eckart and Landau frames, we systematically extract the out-of-equilibrium transport coefficients. We find that the fluid is governed by a negative effective thermal conductivity and a negative heat capacity. This mechanism provides a hydrodynamic interpretation of thermodynamic stability, effectively translating the previously established quasinormal mode damping of the large-$D$ Reissner-Nordström geometry into the language of fluid dynamics.}
\usepackage{float}
\begin{document}
\maketitle
\flushbottom

\section{Introduction}
The large $D$ limit of General Relativity is a highly effective analytical framework for handling nontrivial black hole dynamics. In this limit, the system of coupled nonlinear Einstein equations becomes analytically tractable as the gravitational dynamics of the black hole separates into two widely separated time scales: a long one of order $r_0$ and a short one of order $r_0/D$, where $r_0$ is the characteristic length scale of the black hole, like its Schwarzschild radius~\cite{Emparan:2013moa,Emparan:2014aba,Emparan:2015rva}. As $D$ increases, the separation widens. This phenomenon is also evident in the quasinormal mode spectrum of Schwarzschild and Reissner-Nordström black holes.\footnote{The quasinormal spectrum separates into light and heavy modes. The heavy modes have frequencies $D$ times larger than the light ones, with the imaginary part of the frequencies proportionately large, so these heavy modes die off very quickly after the perturbation, leaving behind only the light modes. The light modes are much fewer in number, so their effective dynamics, although nonlinear, are expected to be much simpler than the full dynamics, which includes both types of modes.} This scale separation arises because the nontrivial metric and gauge field are localised to a thin region of thickness $\mathcal{O}\left({r_0}/{D}\right)$ near the event horizon. This localisation gives rise to the large $D$ membrane paradigm formalism, where the evolution of the black hole is entirely captured by the dynamics of a non-gravitational, codimension-one timelike hypersurface propagating in a non-gravitational background spacetime. The short time scale ($\mathcal{O}({r_0}/{D})$) dynamics decay away rapidly, leaving behind the nonlinear long time scale ($\mathcal{O}({r_0})$) dynamics of the light quasinormal modes. Consequently, the membrane paradigm serves as an effective nonlinear theory of these light modes at long time scales~\cite{Bhattacharyya:2017hpj,Kundu:2018dvx,Bhattacharyya:2015dva,Bhattacharyya:2015fdk,Dandekar:2016fvw}.

Evaluating this effective theory through the lens of hydrodynamic duality is an essential step towards understanding the fluid-gravity correspondence. In the case of asymptotically Anti-de Sitter (AdS) black hole spacetimes, the correspondence is rigorously established~\cite{Bhattacharyya:2007vjd,Dandekar:2017aiv}, and subsequent works have demonstrated the equivalence between the fluid-gravity correspondence, which is written as a derivative expansion in the fluid and the large $D$ membrane-gravity correspondence~\cite{Bhattacharyya:2018iwt,Bhattacharyya:2019mbz,Patra:2019hlq}, where the dual fluid resides on
the asymptotic conformal boundary. Formulating an analogous hydrodynamic duality for asymptotically flat spacetimes presents a distinct physical challenge. Because asymptotically flat spacetimes do not have a timelike boundary, the effective relativistic fluid cannot be projected to spatial infinity; instead, it can be constructed directly upon the dynamical membrane itself.

In this work, we develop a detailed leading order (in $1/D$) hydrodynamic analysis of the effective fluid dual to the charged large $D$ membrane paradigm in an asymptotically flat spacetime. The membrane is formulated as a codimension-one timelike hypersurface propagating in the background flat spacetime~\cite{Bhattacharyya:2015dva,Bhattacharyya:2015fdk,Dandekar:2016fvw}. This embedding allows us to formulate an effective relativistic fluid theory. 

Earlier literature~\cite{Bhattacharyya:2016nhn} established the equivalence between the membrane equations and the conservation of a membrane stress-energy tensor and charge current—identifying its strictly leading order $\mathcal{O}(D)$ structure as an ideal dust and a subleading $\mathcal{O}(1)$ term as shear viscosity of a fluid, which yields the ratio of shear viscosity $\eta$ and entropy density $s$ as $\eta/s = 1/4\pi$. However, a systematic formulation of a dissipative hydrodynamic framework that is consistent with the leading order membrane equations,\footnote {The membrane equations dictate the dynamics of the system, acting as constraint equations that govern membrane quantities such as shape, charge and membrane velocity. Those equations were initially introduced in \cite{Bhattacharyya:2015dva,Bhattacharyya:2015fdk}, building upon the earlier foundational work \cite{Emparan:2013moa,Emparan:2013xia,Emparan:2013oza,Emparan:2014cia,Emparan:2014jca,Emparan:2014aba,Emparan:2015rva}. For the specific case of stationary solutions, alternative derivations of the membrane dynamics can be found in Ref~\cite{Emparan:2015hwa,Suzuki:2015iha,Tanabe:2015isb,Tanabe:2016opw}.} and that identifies the geometric subleading order dissipative corrections to the stress-energy tensor and charge current as other dissipative fluid variables like heat flux and charge diffusion current: this  has notably remained  absent. Specifically, the formulation of out-of-equilibrium thermodynamic variables, the unambiguous definition of distinct fluid frames, and the explicit extraction of the fluid's independent heat flux and charge diffusion coefficients have not yet been established. Without these elements, the equivalence between the dynamics of a charged large $D$ membrane and the dissipative hydrodynamics of a relativistic fluid remains incomplete.

To close this gap, we construct the hydrodynamic theory directly on the $(D-1)$-dimensional membrane world-volume. The mapping is established by mapping the membrane quantities (shape, charge, and membrane velocity) to the local thermodynamic and kinematic fluid variables (temperature, chemical potential, and fluid velocity). We also demonstrate that the leading order geometric dynamics of the membrane is equivalent to the hydrodynamics of a relativistic, charged fluid propagating on the membrane world-volume under the influence of an effective background force. In dissipative relativistic hydrodynamics, the notion of a local rest frame is no longer unique because energy and charge transport cannot, in general, be eliminated simultaneously. This naturally leads to two physically distinguished frame choices: the Eckart frame~\cite{Eckart1940}, where the fluid velocity is aligned with the particle flow, and the Landau frame~\cite{Landau1959}, where it is aligned with the energy flow. Since these frames impose different conditions on the dissipative currents, we perform our analysis in both formulations to obtain a more complete description of the membrane fluid dynamics. Moving beyond global thermodynamic equilibrium, we redefine the local temperature and chemical potential to capture out-of-equilibrium dynamics. Finally, by expressing the Eckart-frame heat flux and the Landau frame charge diffusion current in terms of these redefined variables, we systematically extract the leading-order transport coefficients governing the dissipative membrane fluid in both the Eckart and Landau frames.

Our extraction of the transport coefficients reveals a specific thermodynamic mechanism governing the charged membrane fluid. We establish that the effective fluid is characterised by a negative effective thermal conductivity paired with a negative heat capacity, while negative thermal conductivity inverts the direction of standard heat flow, causing heat to transfer from colder regions to hotter ones. Negative heat capacity reverses the thermal response: an influx of heat lowers the temperature, and an outflux of heat raises it. This mechanism thereby quenches thermal perturbations and helps enforce global thermodynamic equilibrium. Furthermore, we discuss that this mechanism provides a hydrodynamic interpretation of stability, by directly translating the membrane's quasinormal mode analysis of the large-$D$ Reissner-Nordström geometry—in which perturbations naturally damp out, thereby confirming the dynamical validity of the membrane fluid duality.

The paper is organised as follows: Section~\ref{Large D membrane paradigm} revisits the Large D membrane paradigm formalism for a charged black hole. Section~\ref{Fluid dynamical analysis} formulates the Eckart and Landau fluid frames by constructing the corresponding fluid velocities from the membrane data, deriving the fluid equations of motion in both frames in terms of these variables, and demonstrating their equivalence to the leading order membrane equations. In section~\ref{Thermodynamic Stability}, we discuss the thermodynamic stability of the membrane fluid. Section~\ref{Summary and outlook} summarises the analysis with an outlook.
%%%%%%%%%%%%%%%%%%%%%%%%%%%%%%%%%%%%%%%%%%%%%%%%%%%%%%%%%%%%%%%%%%%%%%
\section{Large \texorpdfstring{$D$}{D} membrane paradigm}\label{Large D membrane paradigm}
%%%%%%%%%%%%%%%%%%%%%%%%%%%%%%%%%%%%%%%%%%%%%%%%%%%%%%%%%%%%%%%%%%%%%
The external geometry of an arbitrary dynamical $D$-dimensional charged black hole can be constructed by patching together multiple local boosted charged black brane geometries~\cite{Bhattacharyya:2015fdk}. Each black brane patch is characterised by parameters $r_0$, $u^\mu$ and $Q$. Where $r_0$ represents the event horizon radius, $Q$ represents the charge density of the membrane, and $u^\mu$ represents the boost velocity of a local patch of the corresponding Reissner-Nordström (RN) black hole. For example, for an RN black hole, each patch has a constant charge density field $Q$, boost velocity $u^\mu$, and a constant event horizon radius $r_0$. For an arbitrary black hole, different patches have different values of these parameters.
For general dynamical configurations, the patches are smoothly sewn together such that the collection of their local event horizons seamlessly forms a $(D-1)$-dimensional hypersurface.
\subsection{\texorpdfstring{$D$}{D} tends to infinity}
Consider an RN black hole with mass $m$ and charge $q$ in $D$ spacetime dimensions. The metric and the U(1) gauge potential are given by
\begin{align}
&ds^2= -f(r)dt^2+\frac{dr^2}{f(r)}+r^2d\Omega_{D-2}^2,\label{metric}\\
&A= \frac{q}{(D-3)r^{D-3}\Omega_{D-2}}dt,\label{gauge0}
\end{align}
where
\begin{equation}
f(r)= 1- \frac{16\pi m}{(D-2)\Omega_{D-2}r^{D-3}}+\frac{8\pi q^2}{(D-2)(D-3)\Omega_{D-2}^2r^{2(D-3)}}.
\end{equation}
The parameters $m$ and $q$ are traded for two new parameters $Q$ and $r_0$, which yield a notion of the energy and charge densities on the horizon (membrane) instead of the mass and charge of the black hole. This is necessary to convert the black hole picture to the membrane picture, where the local densities are the fields on the membrane. This reparameterization is given by:
$$m=\frac{(D-2)(1+c_DQ^2)r_0^{D-3}\Omega_{D-2}}{16\pi},$$
and $$q= \frac{(D-3)Qr_0^{D-3}\Omega_{D-2}}{\sqrt{8\pi}},$$
 where  $$c_D\equiv \frac{D-3}{D-2}.$$
This conversion yields:
\begin{align}
&f(r)=\bigg(1-(1+c_DQ^2)\left(\frac{r_0}{r}\right)^{D-3}+c_DQ^2\left(\frac{r_0}{r}\right)^{2(D-3)}\bigg),\label{Expf}\\
&A= \frac{Q}{\sqrt{8\pi}}\left(\frac{r_0}{r}\right)^{D-3}dt.\label{gauge}
\end{align}
Both the nontrivial metric and the gauge potential exhibit extreme localisation in the large $D$ limit~\cite{Emparan:2013moa}. Introducing a radial coordinate $R$ as $r=r_0\left(1+\frac{R}{D-3}\right)$,
as $D\to \infty$, the term $\left(\frac{r_0}{r}\right)^{D-3}$ approaches $\exp(-R)$. Consequently, the nonzero gauge potential and deviation from the flat Minkowski background $(f(r)=1)$ are confined to a "skin thickness" of order $\frac{r_0}{D}$ just outside the event horizon ($r=r_0$). This region is called the "membrane region". As $D$ grows strictly to $\infty$, the membrane region becomes infinitesimally thick and collapses onto the event horizon. Therefore, at any finite distance outside the horizon, the metric is flat, and the gauge potential is zero. Since the spacetime region enclosed by the event horizon is causally disconnected from the outside, we can choose any metric and gauge potential for the inside region. For convenience, we choose it to have a flat metric and a constant gauge potential. This picture is exactly accurate only for the $D=\infty$ case. For a large but finite dimensionality, it is corrected perturbatively in orders of $\frac{1}{D}$. As the leading order of perturbation, we choose a collective coordinate ansatz (Eq. \eqref{Expg_0}, Eq. \eqref{ExpA0}), where in each patch within the skin thickness (which is nonzero for finite $D$), the RN metric and gauge potential exist.

%%%%%%%%%%%%%%%%%%%%%%%%%%%%%%%%%%%%%%%%%%%%%%%%%%%%%%%
\subsection{Membrane embedding and kinematics}
%%%%%%%%%%%%%%%%%%%%%%%%%%%%%%%%%%%%%%%%%%%%%%%%%%%%%
The membrane is defined as a dynamical codimension-one timelike hypersurface propagating in the flat $D$-dimensional Minkowski background. Let the background spacetime be specified with the flat metric $\eta_{MN}$ and coordinates $X^M$. The membrane is defined by the locus of a scalar shape function $\psi(X)$ as $\psi(X) = 1$. The unit normal vector, oriented in the outward direction to the membrane, is given by 
\begin{equation}n_M = \frac{\partial_M \psi}{\sqrt{\eta^{AB}\partial_A \psi \partial_B \psi}}.\end{equation}
The metric restricted to the membrane worldvolume is
\begin{equation}g_{MN} = \eta_{MN} - n_M n_N,\end{equation}
and the extrinsic curvature, which encodes the embedding of the membrane into the flat background, is defined as $K_{MN} = \nabla^{(bg)}_M n_N$, with its trace denoted by $K = g^{MN}K_{MN}$. We define a normalised timelike velocity field $u^\mu$, which always lies in the membrane worldvolume, satisfying $g_{\mu\nu}u^\mu u^\nu = -1$.
%%%%%%%%%%%%%%%%%%%%%%%%%%%%%%%%%%%%%%%%%%%%%%%%%%%%%%%%%%%%%%%%%%%%%%%
\subsection{The \texorpdfstring{$1/D$}{1/D} perturbative expansion}
%%%%%%%%%%%%%%%%%%%%%%%%%%%%%%%%%%%%%%%%%%%%%%%%%%%%%%%%%%%%%%%%%%%%%%%%
To capture the dynamics of the membrane region, we treat $1/D$ as a small expansion parameter and solve the Einstein-Maxwell equation perturbatively. The full dynamical spacetime metric $G_{MN}$ and gauge field $A_M$ are expanded perturbatively around the leading order  large $D$ geometry:
\begin{align}
G_{MN} &= G^{(0)}_{MN} + \frac{1}{D} h_{MN} + \mathcal{O}\left(\frac{1}{D^2}\right),\label{Expgtotal} \\
A_M &= A^{(0)}_M + \frac{1}{D} a_M\label{ExpA:mu} + \mathcal{O}\left(\frac{1}{D^2}\right).
\end{align}
The leading order gauge field is given  as 
\begin{align}\label{ExpA0}
    A^{(0)}_M=\frac{Q\psi^{-D}}{\sqrt{8\pi}}O_M,
\end{align}

and the leading order metric ansatz $G^{(0)}_{MN}$ represents the locally boosted Reissner-Nordström black brane~\cite{Bhattacharyya:2015fdk}. It is constructed out of the membrane data (membrane velocity$u_\mu$, membrane charge density Q, and normal vector $n_\mu$). In the Kerr-Schild form, we can write
\begin{equation}\label{Expg_0}
G^{(0)}_{MN} = \eta_{MN} +\mathcal{F}(\psi, Q) O_M O_N,
\end{equation}
where  $O_M =n_M-u_M$, and $\mathcal{F}(\psi, Q) = (1+Q^2)\psi^{-D} - Q^2 \psi^{-2D}.$
The quantity $\psi$ is defined in the previous subsection, where $\psi=1$ represents the membrane hypersurface, and $\psi>1$ and $\psi<1$ represent the regions "outside" and "inside" the membrane, respectively. For the RN black hole, $\psi=\frac{r}{r_0}$. A detailed analysis of this is done in~\cite{Bhattacharyya:2015fdk}, where it is demonstrated that the substitution of the perturbative ansatz Eq.~\eqref{Expgtotal} and Eq.~\eqref{ExpA:mu} into the $D$-dimensional Einstein-Maxwell equations yields ordinary differential equations in the outward radial coordinate ($R$). The solutions to these equations (which constitute the first-order metric and gauge corrections) remain non-singular if and only if the background membrane data satisfy specific constraint equations given as the leading order membrane equations.
%%%%%%%%%%%%%%%%%%%%%%%%%%%%%%%%%%%%%%%%%%%%%%%%%%%%%%%%%%%%%
\subsection{Membrane stress tensor and charge current}
%%%%%%%%%%%%%%%%%%%%%%%%%%%%%%%%%%%%%%%%%%%%%%%%%%%%%%%%%%%%%%
Those surface constraints, described in the previous subsection, can be written as the conservation equations of the membrane stress-tensor and the membrane charge current. Their explicit computation is performed in~\cite{Bhattacharyya:2016nhn}, where the spacetime curvature (resulting from a deviation from flatness in the metric) and the gauge potential are assumed to be sourced by a stress tensor and a charge current that live on the membrane worldvolume coinciding with the event horizon.
Here we use the effective stress-energy tensor $T_{\mu\nu}$ and the effective charge current $J^\mu$, so that their conservation laws yield leading order membrane equations. Their expressions are as follows:
\begin{equation}\label{EMTensor}
\begin{split}
T_{\mu\nu} &= \left(\frac{1}{8\pi}\right) \left[ \left(\frac{K}{2}\right) (1+Q^2) u_\mu u_\nu + \left(\frac{1-Q^2}{2}\right) K_{\mu\nu} - \left(\frac{\nabla_\mu u_\nu + \nabla_\nu u_\mu}{2}\right) \right. \\
&\quad \left. - (u_\mu v_\nu + u_\nu v_\mu) \right. \bigg]+ \mathcal{O}\left(\frac{1}{D}\right),
\end{split}
\end{equation}
where 
\begin{align}
v_\mu =& Q \nabla_\mu Q + Q^2 (u^\alpha K_{\alpha\mu}) + \left(\frac{2Q^4 - Q^2 - 1}{2}\right) \left(\frac{\nabla_\mu K}{K}\right) - \left(\frac{Q^2 + 2Q^4}{2}\right) (u \cdot \nabla) u_\mu \notag\\ &+ \left(\frac{1+Q^2}{K}\right) \nabla^2 u_\mu,\label{expv}
\end{align}
and
\begin{equation}\label{Chargecurrent}
J^\mu = \left(\frac{Q}{2\sqrt{2\pi}}\right) \left[ K u^\mu -\bigg( \left(\frac{ \nabla_\nu Q}{Q}\right) + (u \cdot \nabla) u_\nu \bigg)P^{\nu\mu} \right]  + \mathcal{O}\left(\frac{1}{D}\right).
\end{equation}
The membrane equations are as follows:
\begin{equation}\label{MembraneEq}
\begin{gathered}
\nabla \cdot u = 0, \\
P^\mu_\nu (u \cdot \nabla) u_\nu = P^\mu_\nu \left( \frac{\nabla^2 u_\nu - (1-Q^2)\nabla_\nu K + K \left(u^\alpha K_{\alpha\nu}\right)}{K(1+Q^2)} \right), \\
u^\nu \nabla_\nu (K Q) = \nabla^2 Q + K Q (u^\alpha K_{\alpha\beta} u^\beta).
\end{gathered}
\end{equation}
All the covariant derivatives and index raising and lowering happen with respect to (D-1) dimensional induced membrane worldvolume metric  $g_{\mu\nu}$. Here, $P_{\mu\nu}$ is the projector, which is perpendicular  to $u^\mu$, defined as $$P_{\mu\nu} = g_{\mu\nu} + u_\mu u_\nu.$$
The first two equations of~\eqref{MembraneEq} come from different components of $\nabla_\mu T^{\mu \nu}=0$ and the last equation comes from the equation $\nabla_\mu J^\mu=0$.
%%%%%%%%%%%%%%%%%%%%%%%%%%%%%%%%%%%%%%%%%%%%%%%%%%%%%%%%%%%%%%%%%%%%%%%%
\section{Fluid dynamical analysis}\label{Fluid dynamical analysis}
%%%%%%%%%%%%%%%%%%%%%%%%%%%%%%%%%%%%%%%%%%%%%%%%%%%%%%%%%%%%%%%%%%%%%%%%
The dynamics of the large $D$ membrane is given by the conservation of its stress-energy tensor and charge current, which are the same as the constraint equations in the Einstein-Maxwell system~\cite{Bhattacharyya:2016nhn}. Using Eq.~\eqref{EMTensor} and Eq.~\eqref{Chargecurrent}, we construct a dual relativistic hydrodynamic framework, termed the ``membrane fluid". This correspondence maps geometric membrane data directly onto macroscopic fluid variables. The quantity $u^\mu$ in Eq.~\eqref{EMTensor} and Eq.~\eqref{Chargecurrent} is a pure geometric quantity called the membrane velocity field. We proceed further by decomposing the conserved quantities into components longitudinal and transverse to $u^\mu$.
From Eq.~\eqref{Chargecurrent}, the membrane charge current can be expressed as
\begin{align}\label{Chargecurrent2}
   J^\mu =L u^\mu +M^{\mu},
\end{align}
where
\begin{align}\label{EqL&M}
&L=\frac{QK}{2\sqrt{2\pi}},
&&M^\mu=\frac{Q}{2\sqrt{2\pi}}  \left(- \frac{\nabla_\nu Q}{Q} P^{\mu\nu}  - u \cdot \nabla u^\mu\right). 
\end{align}
Here, $M^\mu$ is perpendicular to $u_\mu$ satisfying,
$u_\mu M^\mu =0.$ The stress-energy tensor from Eq.~\eqref{EMTensor} can be written  as 
\begin{equation}\label{EMAppendix}
T_{\mu\nu} = s_1 u_\mu u_\nu -\frac{ v_\mu}{8\pi} u_\nu -\frac{ v_\nu}{8\pi} u_\mu + w_{\mu\nu}+\mathcal{O}(1/D),
\end{equation}
\text{ where }
\begin{align}\label{Exps1w}
&s_1 = \frac{K(1+Q^2)}{16\pi},
&& w_{\mu\nu} =\frac{1}{8\pi}\left[ \left(\frac{1-Q^2}{2}\right) K_{\mu\nu} - \frac{\nabla_\mu u_\nu + \nabla_\nu u_\mu}{2}\right].
\end{align}
Decomposing different elements of $T_{\mu\nu}$ along and orthogonal to $u_\mu$ as 
\begin{align}\label{EMTensor2}
T_{\mu\nu} &= \epsilon u_\mu u_\nu + l_\mu u_\nu + l_\nu u_\mu + r_{\mu\nu}+\mathcal{O}(1/D) ,
\end{align}
where up to $\mathcal{O}(1)$ the following relations are obeyed: $$T_{\mu\nu}u^{\nu}u^{\mu}=\epsilon,\quad T_{\mu\nu}u^{\nu}=-l_{\mu}-\epsilon u_{\mu}
, \quad u^{\mu}l_{\mu}=0,\quad r_{\mu\nu}u^{\nu}=0.$$
The expressions of $\epsilon, l_\mu, r_{\mu\nu}$ are given in Table~\ref{table1}. For an explicit calculation of the quantities listed in Table~\ref{table1}, see Appendix~\ref{A1}.
\setcounter{table}{0} 
\begin{table}[H]
    \centering
    \caption{Geometric Quantities}
    \renewcommand{\arraystretch}{2.2}
    \begin{tabular}{|l|l|}
    \hline
    \textbf{Scalar} & 
    $\begin{aligned}
    \epsilon &= \frac{K}{16\pi}(1+Q^2) + \frac{1}{8\pi} \bigg[ \frac{(3 Q^2+1)}{2}  u^\alpha u^\beta K_{\alpha\beta} + 2Q u \cdot \nabla Q \\
      &\quad + \frac{(2Q^4 - Q^2 - 1)}{K} u \cdot \nabla K \bigg]
    \end{aligned}$ \\ \hline
    \textbf{Vector} & 
    $\begin{aligned}
    l_\mu &= -\frac{1}{8\pi}\bigg[ Q \nabla^b Q +\frac{1+Q^2}{2} (u_\alpha K^{\alpha b}) + \left(\frac{2Q^4 - Q^2 - 1}{2}\right) \left(\frac{\nabla^b K}{K}\right) \\
          &\quad + \left(\frac{1+Q^2}{K}\right) \nabla^2 u^b - \left(\frac{1+Q^2 + 2Q^4}{2}\right) (u \cdot \nabla) u^b\bigg]P_{b \mu}
    \end{aligned}$ \\ \hline
    \textbf{Tensor} & 
    $\begin{aligned}
    r_{\mu\nu} &= \frac{1}{16\pi} \left[ (1-Q^2) K^{\alpha\beta} - (\nabla^\alpha u^\beta + \nabla^\beta u^\alpha) \right] P_{\alpha\mu} P_{\beta\nu}
    \end{aligned}$ \\ \hline
    \end{tabular}
    \label{table1}
\end{table}
\subsection{Choice of hydrodynamic frame}\label{Choice of hydrodynamic frame}
Formulating the hydrodynamics of a viscous fluid requires specifying a hydrodynamic frame to resolve the ambiguity in defining the out-of-equilibrium quantities. There are two standard conventions: the Eckart frame, in which the fluid velocity is aligned with the conserved-charge current, and the Landau frame, which aligns it with the energy flow~\cite{Eckart1940,Landau1959}. In this section, we analyse the membrane fluid in both frames. By constructing the corresponding velocities, we determine the tensor structure of the dissipative quantities in terms of geometric quantities, which is a necessary prerequisite for computing the transport coefficients of the membrane fluid.
\subsubsection{Eckart frame}\label{Eckart Frame}
\noindent {In the Eckart frame, the hydrodynamic velocity field is defined by the particle (or charge) flow. This choice implies that the spatial charge diffusion current vanishes. Consequently, the Eckart velocity $U^\mu$ is strictly proportional to $J^\mu$. Identifying the membrane charge current as the charge current of the dual relativistic fluid and imposing the standard normalisation condition $U^\mu U_\mu = -1$,  the velocity field is defined  as 
\begin{align}\label{Eckartvelocity}
    U^\mu &= \frac{J^\mu}{\sqrt{-J^2}} = \frac{L u^\mu + M^\mu}{\sqrt{L^2 - M^2}}=u^\mu +\frac{M^\mu }{L}+\mathcal{O}(\frac{1}{D^2}),
\end{align}
where $J^\mu$ can be expressed  as 
\begin{align}\label{EckartChargeCurrent}
    J^\mu = n U^\mu,
\end{align} 
 with, \begin{align}
      n & = \sqrt{L^2 - M^2}=  \frac{Q K}{2\sqrt{2\pi}}  + \mathcal{O}(1/D).\label{Expn}
 \end{align}

\noindent After explicit substitution of $u^\mu$ in terms of $U^\mu$ from Eq.~\eqref{Eckartvelocity}, Eq.~\eqref{EMTensor2} becomes,
\begin{align}
    T_{\mu\nu} &= \epsilon U_\mu U_\nu  + q_\mu U_\nu + q_\nu U_\mu + r_{\mu\nu} + \mathcal{O}(1/D) ,\label{EckartEMTensorGeo}
\end{align}
 where 
\begin{align}
q_\mu = &\left( l_\mu - \frac{\epsilon}{L} M_\mu \right), \notag\\ 
     &= -\bigg( \frac{Q^2-1}{2Q} {\nabla}^\alpha Q + \frac{(1+Q^2)}{K} {\nabla}^2 u^\alpha  + \frac{(2Q^4 - Q^2 - 1)}{2K} {\nabla}^\alpha K 
    + \frac{(Q^2+1)}{2} u_\beta K^{\beta \alpha}\nonumber\\& -(1+Q^2+Q^4) u \cdot \nabla u^\alpha\bigg) \frac{P_{\mu\alpha}}{8\pi}.\nonumber
\end{align}
By substituting the explicit form of $u_\beta K^{\beta \alpha}$ from membrane Eq.~\eqref{MembraneEq} in the above expression, we get the following basis expansion:
\begin{align}\label{HeatFluxGeometric}
q_{\mu} &= \bigg( \frac{1-Q^2}{Q} {\nabla}^\alpha Q -\frac{(Q^2+1)}{K} {\nabla}^2 u^\alpha  + \frac{Q^2(1-Q^2)}{K} {\nabla}^\alpha K+(Q^4+1) u.\nabla u^\alpha 
     \bigg) \frac{P_{\mu\alpha}}{16\pi}.
\end{align}
Now we define tensor $T^{\text{fluid}}_{\mu \nu}$ as stress-energy tensor of dual relativistic fluid  as 
\begin{align}\label{Eck.Fluid.EM}
    T^{\text{fluid}}_{\mu\nu} &= \epsilon U_\mu U_\nu + q_\mu U_\nu + q_\nu U_\mu + \tau_{\mu\nu} + \mathcal{O}(1/D),
\end{align}
where $\epsilon$,$n$, $q_{\mu}$,$\tau_{\mu\nu}$ are energy density, charge density, heat flux and shear stress tensor of the fluid, respectively. The expression of shear stress tensor $\tau_{\mu\nu}$ is defined  as }
\begin{align}\label{Eck.ShearStress}
\tau_{\mu \nu} &= -\frac{1}{16\pi} (\nabla^\alpha U^\beta + \nabla^\beta U^\alpha) P_{\alpha\mu} P_{\beta\nu}+ \mathcal{O}(1/D),
\end{align}
 where 
$$P_{ab}=g_{ab}+U_a U_b +\mathcal{O}(1/D),$$ 
and up to $\mathcal{O}(1)$ the following relations are satisfied:
$$T^{\text{fluid}}_{\mu\nu}U^{\nu}U^{\mu}= \epsilon,\quad T^{\text{fluid}}_{\mu\nu}U^{\nu}=-q_{\mu}-\epsilon U_{\mu},\quad U^{\mu}q_{\mu}=0,\quad \tau_{\mu\nu}U^{\nu}=0.$$
Under this identification, the geometric constraint equations governing the membrane shape and charge ($\nabla_{\mu}T^{\mu\nu}=0$ and $\nabla_{\mu}J^{\mu}=0$), can be interpreted as the energy-momentum conservation and charge conservation equations of a forced fluid, with a force density $f^\nu$, generating from the background geometry (the extrinsic curvature tensor of membrane). The equation of motion of the fluid takes the form
\begin{align}\label{FluidEMEq}
    \nabla_{\mu}T^{(\text{fluid})\mu\nu}=f^\nu,
    \end{align}
    with,
    \begin{align}\label{Forceform}
    f^\nu=\frac{1}{16\pi}\nabla_{\mu}\left[(Q^2-1) K^{\alpha\beta}P_{\alpha}^{\mu} P_{\beta}^{\nu}\right]. 
    \end{align}
Notably, the continuity equation remains the same as 
    \begin{align}\label{FluidChargeEq}
        \nabla_{\mu}J^{\mu}=0.
    \end{align}
Relativistic hydrodynamics usually assumes a system where energy and momentum are conserved. The membrane fluid, on the other hand, is a hydrodynamic system driven by a background force. This contrasts with the standard AdS fluid-gravity correspondence, where the fluid exists on a geometrically rigid asymptotic boundary. The large $D$ membrane is a dynamical hypersurface embedded in a background flat spacetime. This embedding geometry generates the effective force density $f^{\nu}$ from the extrinsic curvature $K^{\alpha\beta}$ (Eq.~\eqref{Forceform}). This force has direct consequences on the leading order transport coefficients; it dictates the fluid's acceleration, which leads to the emergence of independent transport coefficients, as noted in Sections~\ref{Analysis in Eckart frame} and \ref{Analysis in Landau frame}.

The Gauss-Codazzi equations link the extrinsic curvature to the intrinsic Ricci tensor $R_{\mu\nu}$ of the membrane world-volume. For a flat background spacetime, the intrinsic Ricci tensor of the embedded hypersurface is exactly $$R_{\mu\nu} = K K_{\mu\nu} - K^\alpha_\mu K_{\alpha\nu}.$$ 
This identity raises the possibility that the background force could be recast and alternatively interpreted as the consequence of a nonminimal coupling of the fluid with the induced intrinsic spacetime curvature of the membrane world-volume. This intrinsic curvature could be seen as a gravitational field by the fluid from the membrane world-volume point of view. A rigorous analysis of this alternative geometric interpretation is beyond the scope of this paper and is deferred to future work.

It is important to note that to obtain the complete leading order membrane equations from the requirement that the stress-energy tensor is conserved, one must incorporate some specific terms from the subleading order in $(1/D)$ into the stress-energy tensor.\footnote{Those specific terms in the stress-energy tensor produce an extra factor of $D$ while taking the divergence to compute the membrane equations. Hence, they contribute to the leading order membrane equations.} However, the full subleading order stress-energy tensor is not needed for the present leading-order analysis, though it would be required for deriving the subleading order membrane equations. In Ref.~\cite{Biswas:2019xip}, the exact computation of the full subleading order stress tensor is performed for the uncharged membrane paradigm, where the corresponding pressure part arises at strict subleading order in the stress-energy tensor. For this reason, it does not enter the leading order membrane equations. So for computing the leading order membrane equation in $\mathcal{O}(1/D)$, the corresponding membrane fluid behaves as a pressure-suppressed viscous fluid.
 %%%%%%%%%%%%%%%%%%%%%%%%%%%%%%%%%%%%%%%%%%%%%%%%%%%%%
\subsubsection{Landau frame}\label{Landau Frame}
%%%%%%%%%%%%%%%%%%%%%%%%%%%%%%%%%%%%%%%%%%%%%%%%%%%%%
The Landau frame aligns the fluid velocity exclusively with the energy flow. Consequently, the velocity field is fixed as the normalised, timelike eigenvector of the stress-energy tensor. We implement this frame choice by defining the transformed velocity vector of fluid as 
\begin{equation}\label{LandauVelocity}
   \tilde{U}_\mu = U_\mu  + \frac{q_\mu}{\epsilon }.
\end{equation}
After explicit substitution \footnote{For the explicit calculation to prove the validity of this substitution up to $\mathcal{O}$(1), see Appendix~\ref{A0}.} of $U_\mu$ in terms of $\tilde{U}_\mu$ from Eq.~\eqref{LandauVelocity}, in Eq.~\eqref{Eck.Fluid.EM} and Eq.~\eqref{EckartChargeCurrent}, the expression of the stress-energy tensor and charge current in the Landau frame takes the form as 
\begin{align}
    T^{\text{fluid}}_{\mu\nu} &= \epsilon \tilde{U}_\mu \tilde{U}_\nu + \tau_{\mu\nu} + \mathcal{O}\left(\frac{1}{D}\right),\label{LandauFluidEM} \\
    J_\mu &= n \left(\tilde{U}_\mu -\frac{q_\mu}{\epsilon} \right) +\mathcal{O}\left(\frac{1}{D}\right),\nonumber \\
          &= n \tilde{U}_\mu + N_\mu +\mathcal{O}\left(\frac{1}{D}\right),\label{LandauChargeCurrent}
\end{align}
 where the following relations are obeyed up to $\mathcal{O}(1)$:
\begin{align*}
    &\tilde{U}^\mu \tilde{U}_\mu = -1, && T^{\text{fluid}}_{\mu\nu}\tilde{U}^\mu \tilde{U}^\nu = \epsilon,  \\
    &T^{\text{fluid}}_{\mu\nu} \tilde{U}^\nu = -\epsilon\tilde{U}_\mu,   &&\tau_{\mu\nu} \tilde{U}^\nu= 0, &&& N^\mu\tilde{U}_\mu = 0.
\end{align*}
With the definition of charge diffusion current $N_\mu$, shear-stress tensor $\tau_{\mu \nu}$ and projector $P_{ab}$ in the Landau frame  as 
\begin{align}
&P_{ab}=g_{ab}+\tilde{U}_a \tilde{U}_b +\mathcal{O}(1/D),\notag\\
&\tau_{\mu \nu} = -\frac{1}{16\pi} (\nabla^\alpha \tilde{U}^\beta + \nabla^\beta \tilde{U}^\alpha) P_{\alpha\mu} P_{\beta\nu}+\mathcal{O}(1/D),\label{LandauShearStress}\\
&N_\mu = -\frac{n q_\mu}{\epsilon }.\label{ExpN_a}
\end{align}
The energy density $\epsilon$ and charge density $n$ coincide up to the first subleading order across both frames. Furthermore, the dynamics of the system continue to be governed by the forced equations of motion for the stress-energy tensor and charge current, given by Eq.~\eqref{FluidEMEq} and Eq.~\eqref{FluidChargeEq}, respectively.
Explicit substitution of Eq.~\eqref{HeatFluxGeometric} in Eq.~\eqref{ExpN_a} yields
\begin{align}\label{NumberDiffCurrGeo}
N_\mu &= -\frac{Q}{2(1+Q^2)\sqrt{2\pi}}\bigg( \frac{1-Q^2}{Q} {\nabla}^\alpha Q -\frac{(Q^2+1)}{K} {\nabla}^2 u^\alpha  + \frac{Q^2(1-Q^2)}{K} {\nabla}^\alpha K\\&\nonumber
+(Q^4+1) u.\nabla u^\alpha 
     \bigg)  P_{\mu\alpha}.
\end{align}
%%%%%%%%%%%%%%%%%%%%%%%%%%%%%%%%%%%%%%%%%%%%%%%%%%%%%%%%%%%%%%%%%%%%
\subsection{Global thermodynamic equilibrium}\label{Global Thermodynamic Equilibrium}
%%%%%%%%%%%%%%%%%%%%%%%%%%%%%%%%%%%%%%%%%%%%%%%%%%%%%%%%%%%%%%%%%%%%
Global thermodynamic equilibrium is a macroscopic state with thermodynamic potentials (temperature, pressure, chemical potential) having the same value everywhere, ensuring no net flux of energy or matter. In this state, the fluid maintains a uniform velocity field. Consequently, all dissipative contributions to the stress-energy tensor and charge current vanish, and they assume an ideal fluid form. However, a critical distinction must be drawn: an ideal fluid is defined by the absence of transport coefficients, whereas a fluid in global thermodynamic equilibrium is still fundamentally dissipative. The dissipation terms vanish because the spatial gradients of the macroscopic quantities, which act as the driving thermodynamic forces, are zero. To realise this equilibrium state of a membrane fluid, we evaluate the membrane dual to a spherically symmetric RN black hole, which is defined (for details see~\cite{Bhattacharyya:2015fdk}) as:
\begin{align*}
 &r=r_0,&&Q=\text{Constant},&&&u^\mu=\text{Constant},\\ &u^\mu K_{\nu\mu} u^\nu=0,&&u^\mu K_{\nu\mu} P^\nu_\alpha=0,&&& K_{\mu\nu}P^\mu_\alpha P^\nu_\beta=g_{\alpha\beta},   
\end{align*}
where $g_{\mu\nu}$ is a metric induced in $(D-1)$-dimensional membrane world volume. 
The stress-energy tensor and charge current of the corresponding membrane fluid become
\begin{equation}\label{EMGlobal}
    T^{\text{fluid}}_{\mu\nu} =  \left(\frac{K}{16\pi}\right) (1+Q^2) u_\mu u_\nu ,
\end{equation}
and
\begin{equation}\label{CharCurrGlobal}
    J^\mu =\frac{QK}{2\sqrt{2\pi}}u^\mu,
\end{equation}
respectively.
Using the above expressions, the RN limit can be interpreted as the global thermodynamic equilibrium state of the membrane fluid, in which the background force vanishes, and $u^\mu$ serves as the fluid's uniform velocity field. The consistency of this duality is verified by Eq.~\eqref{Eckartvelocity} and Eq.~\eqref{LandauVelocity}, which show that the geometric expressions for the fluid's velocity in both the Eckart and Landau frames converge to $u^\mu$ in the RN limit. As thermodynamic variables are rigorously well-defined only in the equilibrium state, the energy density $\epsilon$ and charge density $n$ of the membrane fluid can be unambiguously extracted from Eq.~\eqref{EMGlobal} and Eq.~\eqref{CharCurrGlobal}  as 
\begin{align}
    &\epsilon=\frac{K(1+Q^2)}{16\pi},\label{Exp:epsilon} \\
    &n=\frac{QK}{2\sqrt{2\pi}}\label{Exp:n:Geo}.
\end{align}
The Bekenstein-Hawking entropy formula~\cite{Bekenstein1973,Hawking1975} states that black hole entropy is proportional to its horizon area. In the large $D$ membrane paradigm~\cite{Bhattacharyya:2015dva,Bhattacharyya:2015fdk}, this horizon area is dual to the spatial volume of the effective thin hypersurface of the membrane. Consequently, mapping this geometry to the dual relativistic fluid, the total entropy of the fluid is proportional to its spatial volume of where it lives. So the total entropy of the fluid yields $S = V_{\text{fluid}}/4$. It follows that the fundamental entropy density of membrane fluid is
\begin{equation}\label{Exps}
    s = 1/4.
\end{equation}
We explicitly adopt this exact identification for the formulation throughout this fluid analysis.
To derive the expression of temperature $T$ and chemical potential $\mu$ of the fluid, we apply the first law of thermodynamics to a quasi-static transition between global equilibrium states of the fluid. This corresponds to a transition between RN black hole configurations with varying parameters, specifically from $(Q_0, r_0)$ to $(Q_0+dQ_0, r_0+dr_0)$. To formulate the first law in terms of geometric quantities, we define the total energy $E$, charge $N$, and entropy $S$ by multiplying the fluid volume (which is the area of a codimension-one hypersurface of the membrane) with their respective geometric densities ($e$, $n$, and $s$). This substitution yields the following geometric forms:
\begin{align}
    &E= \frac{K(1+Q^2)}{16\pi} r^{D-2}_0 \Omega_{D-2}=\frac{(D-2)(1+Q^2)}{16\pi} r^{D-3}_0 \Omega_{D-2},\label{ExpE}\\
         &N=\frac{KQ}{2\sqrt{2\pi}} r^{D-2}_0 \Omega_{D-2}=\frac{(D-2)Q}{2\sqrt{2\pi}} r^{D-3}_0 \Omega_{D-2} ,\label{ExpN}
        \\
    &S=\frac{ r^{D-2}_0 \Omega_{D-2}}{4}.\label{ExpS}
\end{align}
Here, $K=\frac{(D-2)}{r_0}$, and $r_0$ is the radius of the event horizon. Differentiation of Eqs.~\eqref{ExpE}-\eqref{ExpS} yields
\begin{align*}
dE &= \frac{(D-2)(1+Q^2)}{16\pi} (D-3) r_0^{(D-4)} \Omega_{D-2} dr_0
 + 2\frac{(D-2)}{16\pi} \Omega_{D-2} r_0^{(D-3)} Q dQ ,\\[10pt]
dN &= \frac{(D-2)Q}{2\sqrt{2\pi}} (D-3) r_0^{(D-4)} \Omega_{D-2} dr_0 + \frac{(D-2)}{2\sqrt{2\pi}} r_0^{(D-3)} \Omega_{D-2} dQ, \\[10pt]
dS &= \frac{(D-2)}{4} r_0^{(D-3)} \Omega_{D-2} dr_0.
\end{align*}
Inserting them in the first law,
$dE = T dS + \mu dN,$ we obtain
\begin{align*}
&(D-3) r_0^{(D-4)}  \left[ \frac{(1+Q^2)}{16\pi} - \frac{\mu Q}{2\sqrt{2\pi}} - \frac{T r_0}{4(D-3)} \right]dr_0 +  \left[ \frac{Q}{8\pi} - \frac{\mu}{2\sqrt{2\pi}} \right]  r_0^{(D-3)}dQ = 0.
\end{align*}
Both the coefficient of $dr_0$ and $dQ$ should be zero. By solving them, we get
\begin{align}
&\mu = \frac{Q}{2\sqrt{2\pi}},\label{ChemicalPot} \\
&T = \frac{(D-3)(1-Q^2)}{4\pi r_0} = \frac{K}{4\pi} (1-Q^2) +\mathcal{O}(1).\label{Temp}
\end{align}
It is important to emphasise that these expressions for chemical potential and temperature for our membrane fluid match the leading order thermodynamic quantities of the corresponding black hole. As discussed in Ref.~\cite{Bhattacharyya:2016nhn}, the chemical potential of black hole was derived by taking the difference between the gauge field evaluated at the event horizon and at asymptotic infinity, and the temperature was calculated via analytic continuation to Euclidean signature and recognising the periodicity of the time cycle, which keeps the solution regular at the event horizon. The fluid-dynamical treatment of membranes reproduces the known black hole result.
Our explicit extraction of $\epsilon, n, \mu, s, T$ enables us to verify that at leading order in $1/D$, the membrane fluid obeys - {\it the thermodynamic Euler relation}.
Specifically Eqs.~\eqref{Exp:epsilon},~\eqref{Exp:n:Geo},~\eqref{Exps},~\eqref{ChemicalPot}, and~\eqref{Temp} satisfy,
\begin{align}\label{Euiler}
    \epsilon=\mu n+Ts.
\end{align}
We refer to section~\ref{A4} for details. Using Eqs.~\eqref{ChemicalPot},~\eqref{Temp},~\eqref{Fluid:epsilon}, and~\eqref{Fluid:n}, the energy and charge densities of the membrane fluid can be expressed in terms of the temperature and chemical potential at global thermodynamic equilibrium as 
\begin{align}
    &\epsilon(T,\mu)=\frac{T(1+8\pi\mu^2)}{4(1-8\pi\mu^2)},\label{Fluid:epsilon}\\ &n(T,\mu)=\frac{4\pi\mu T}{1-8\pi\mu^2}\label{Fluid:n}.
\end{align}
%%%%%%%%%%%%%%%%%%%%%%%%%%%%%%%%%%%%%%%%%%%%%%%%%%%%%%%%%%%%%%%
\subsection{Local thermodynamic equilibrium}\label{Local Thermodynamic Equilibrium}
%%%%%%%%%%%%%%%%%%%%%%%%%%%%%%%%%%%%%%%%%%%%%%%%%%%%%%%%%%%%%%%
When the fluid is out of equilibrium, temperature and thermodynamic potentials develop spacetime gradients. Those gradients act as thermodynamic forces that accelerate the fluid element and drive dissipative fluxes, such as the number (or charge) current and the heat current. Here, we need to introduce the notion of local thermodynamic equilibrium, where we define local temperature, local chemical potential and local velocity field. Furthermore, there can be different choices (different frames) of those definitions.\footnote{ The definitions of thermodynamic quantities at local thermodynamic equilibrium are ambiguous, and this ambiguity is resolved by choosing different frames, but the quantities that are non-ambiguous in all of the frames are the stress-energy tensor and the charge current.} At the equilibrium limit, they should agree with the equilibrium value of those quantities.

 Similar to Ref.~\cite{Kovtun:2019hdm},  we redefine our out-of-equilibrium local temperature $T'(x)$ and local chemical potential $\mu'(x)$ so that the form of equilibrium energy density and charge density is identical with the out-of-equilibrium form of them, only the equilibrium temperature and chemical potential ($T$ and $\mu$) in those expressions, (as described by Eqs.~\eqref{Fluid:epsilon}, and~\eqref{Fluid:n} respectively) are replaced by the redefined out-of-equilibrium temperature and chemical potential ($T'(x)$ and $\mu'(x)$). This ensures that the following equations are satisfied:
\begin{align}
    &\epsilon(T',\mu')=\frac{K(1+Q^2)}{16\pi} + \frac{1}{8\pi} \bigg[  \frac{(3Q^2+1)}{2} u^\alpha u^\beta K_{\alpha\beta} + 2Q u \cdot \nabla Q + \frac{(2Q^4 - Q^2 - 1)}{K} u \cdot \nabla K   \bigg],\label{Redefinition1} \\
    &n(T',\mu')=\frac{QK}{2\sqrt{2\pi}}.\label{Redefinition2}
\end{align}
 We define the out-of-equilibrium temperature and chemical potential  as 
 \begin{align}
     T'(x)=T(x)+\delta T,\label{ExpT'}\\ \mu'(x)=\mu(x)+\delta \mu,\label{Exp:mu'}
 \end{align}
 where $\delta T$ and $\delta \mu$ are corrections in temperature and chemical potential respectively. As the out-of-equilibrium correction in Eq.~\eqref{Redefinition1} is of subleading order in $1/D$, we assume $\delta T$ and $\delta \mu$ to be subleading order corrections to $T$ and $\mu$.  After explicit calculation (see Appendix~\ref{A2}), we find the corrections as
\begin{align}
    &\delta T= \frac{1+Q^2}{2\pi(1-Q^2)} \bigg[ \frac{(3Q^2+1)}{2}  u^\alpha u^\beta K_{\alpha\beta} + 2Q u \cdot \nabla Q + \frac{(2Q^4 - Q^2 - 1)}{K} u \cdot \nabla K   \bigg],\label{Exp:delta:T}\\ 
    &\delta\mu=\frac{-Q}{\sqrt{2\pi}K(1-Q^2)} \bigg[  \frac{(3Q^2+1)}{2}  u^\alpha u^\beta K_{\alpha\beta} + 2Q u \cdot \nabla Q + \frac{(2Q^4 - Q^2 - 1)}{K} u \cdot \nabla K   \bigg].\label{Exp:delta:mu}
\end{align}  
At the equilibrium limit (for RN limit), Eq.~\eqref{Exp:delta:T} and Eq.~\eqref{Exp:delta:mu} vanish, and the out-of-equilibrium temperature and chemical potential  ($T$', $\mu'$) become the equilibrium temperature and chemical potential ($T$ and $\mu$).
%%%%%%%%%%%%%%%%%%%%%%%%%%%%%%%%%%%%%%%%%%%%%%%%%%%%%%
\subsection{Breakdown of the Gibbs-Duhem relation}
\label{A4}
%%%%%%%%%%%%%%%%%%%%%%%%%%%%%%%%%%%%%%%%%%%%%%%%%%%%%%
In this section, we demonstrate that while the thermodynamic Euler relation is satisfied at the leading order in the large $D$ limit, the Gibbs-Duhem relation does not hold exactly for this membrane fluid, at the leading order. From section~\ref{Global Thermodynamic Equilibrium}, the intensive variables and the corresponding densities are given by,
\begin{align}
    \mu &= \frac{Q}{2\sqrt{2\pi}}, \label{eq:mu_app} \\[5pt]
    T &= \frac{(D-3)(1-Q^2)}{4\pi r_0}, \label{eq:T_app} \\[5pt]
    n &= \frac{Q(D-2)}{2\sqrt{2\pi} r_0}, \label{eq:n_app} \\[5pt]
    s &= \frac{1}{4}. \label{eq:s_app}
\end{align}
Further, we obtain the combination $Ts + \mu n$ as follows,
\begin{align}
    Ts + \mu n &= \left( \frac{(D-3)(1-Q^2)}{4\pi r_0} \right) \frac{1}{4} + \left( \frac{Q}{2\sqrt{2\pi}} \right) \left( \frac{Q(D-2)}{2\sqrt{2\pi} r_0} \right), \nonumber \\[5pt]
    &= \frac{(D-3)(1-Q^2)}{16\pi r_0} + \frac{Q^2(D-2)}{8\pi r_0},\\
    &= \frac{D - 3 - D Q^2 + 3 Q^2 + 2 D Q^2 - 4 Q^2}{16\pi r_0}, \nonumber \\[5pt]
    &= \frac{D - 3 + Q^2(D - 1)}{16\pi r_0}. \label{eq:ts_mun}
\end{align}
The energy density of the membrane fluid, defined in terms of geometric quantities, is given by,
\begin{align}
    \epsilon &= \frac{K(1+Q^2)}{16\pi} = \frac{D-2}{16\pi r_0}(1+Q^2), \nonumber \\[5pt]
    &= \frac{D - 2 + Q^2(D - 2)}{16\pi r_0}. \label{eq:epsilon_app}
\end{align}
Subtracting Eq.~\eqref{eq:ts_mun} from Eq.~\eqref{eq:epsilon_app}, we find the exact discrepancy:
\begin{align}
    \epsilon - (Ts + \mu n) &= \frac{(D - 2) - (D - 3) + Q^2(D - 2) - Q^2(D - 1)}{16\pi r_0}, \nonumber \\[5pt]
    &= \frac{1 - Q^2}{16\pi r_0}. \label{eq:euler_mismatch}
\end{align}
Eq.~\eqref{eq:euler_mismatch} proves that $\epsilon = Ts + \mu n$ is satisfied at the leading order in $1/D$, since both $\epsilon$ and $Ts + \mu n$ are $\mathcal{O}(D)$, while their exact difference is strictly $\mathcal{O}(1)$. We can write the first law of thermodynamics for the membrane fluid as
\begin{align*}
    &dE=T dS + \mu dN,\\&
     d(\epsilon V)=T d(sV) + \mu d(nV),\\&
      V d\epsilon + \epsilon dV=Ts dV + TV ds+\mu n dV + \mu V dn,\\& 
      V (d\epsilon - T ds - \mu dn) = -(\epsilon - Ts - \mu n) dV.
\end{align*}
The explicit variation of the total volume where the fluid lives (which is the area of the membrane hypersurface $V = r_0^{D-2}\Omega_{D-2}$) is given by, $$dV = (D-2)r_0^{D-3}\Omega_{D-2}dr_0 = \frac{D-2}{r_0} V dr_0.$$ Substituting this along with Eq.~\eqref{eq:euler_mismatch} we obtain
\begin{align*}
    V (d\epsilon - T ds - \mu dn) = - V \left( \frac{1-Q^2}{16\pi r_0} \right) \left( \frac{D-2}{r_0} dr_0 \right),\nonumber\\
    d\epsilon - T ds - \mu dn = - \frac{(D-2)(1-Q^2)}{16\pi r_0^2} dr_0 =\mathcal{O}(D).
\end{align*}
Since for the membrane fluid $s$ is constant, we obtain,
\begin{align}\label{FirstLaw}
    d\epsilon  - \mu dn \sim\mathcal{O}(D).
\end{align}
Substitution of $\epsilon$ using Eq.~\eqref{eq:euler_mismatch} in the above equation yields the following:
\begin{align}\label{GibbsDuhemMissmatch}
    sdT+nd\mu\sim\mathcal{O}(D).
\end{align}
Eq.~\eqref{GibbsDuhemMissmatch} represents that the equilibrium Gibbs-Duhem relation is not satisfied, which relates the transition between two global thermodynamic equilibria. Now, we can check the local Gibbs-Duhem relation by explicitly substituting the redefined fluid variables (as discussed in section~\ref{Local Thermodynamic Equilibrium}) in the following combination:
\begin{align*}
    sdT'(x)+n(T',\mu')d\mu'(x)&=sdT(x)+n(T,\mu)d\mu(x)+\mathcal{O}(1),\\&=\frac{1}{4}d\bigg(\frac{K(x)(1-Q(x)^2)}{4\pi }\bigg)+\frac{Q(x)K(x)}{2\sqrt{2\pi}}d\bigg(\frac{Q(x)}{2\sqrt{2\pi}}\bigg)+\mathcal{O}(1),\\&=\frac{(1-Q(x)^2)dK(x)}{16\pi}+\mathcal{O}(1),\\&=-\frac{(D-2)(1-Q(x)^2)dr_0(x)}{16\pi r_0(x)^2}+\mathcal{O}(1),\\&=\mathcal{O}(D).
\end{align*}
The above equation shows the breakdown of the local Gibbs-Duhem relation at leading order in $D$.
The mathematical failure of the equilibrium Gibbs-Duhem relation at leading order in $1/D$ originates directly from the dimensional scaling of the membrane's volume element. Because the spatial volume scales as $V \propto r_0^{D-2}$, taking its differential introduces an extra factor of $\mathcal{O}(D)$. Therefore, even though the Euler relation is perfectly valid at the leading order $\mathcal{O}(D)$ with subleading terms safely suppressed to $\mathcal{O}(1)$, the large $D$ expansion mathematically escalates this $\mathcal{O}(1)$ correction to leading order when the volume differential is taken. This dimensional scaling mathematically forces the Gibbs-Duhem relation to break down precisely at the leading $\mathcal{O}(D)$ order. Further investigation also proves the local breakdown. This local breakdown has fundamental consequences for the hydrodynamic framework. In sections~\ref{Analysis in Eckart frame} and~\ref{Analysis in Landau frame}, we have explained in detail how this specific failure gives rise to the independent transport coefficients of the fluid.

%%%%%%%%%%%%%%%%%%%%%%%%%%%%%%%%%%%%%%%%%%%%%%%%%%%%%%%%%%%%%%%
\subsection{Hydrodynamic equations and constitutive relations}\label{Hydrodynamic equations and constitutive relations}
%%%%%%%%%%%%%%%%%%%%%%%%%%%%%%%%%%%%%%%%%%%%%%%%%%%%%%%%%%%%%%%
Following our definition of the out-of-equilibrium temperature $T'$ and chemical potential $\mu'$ in the previous subsection, and the fluid velocity field in subsection~\ref{Choice of hydrodynamic frame}, here, we incorporate those variables in the expression of the stress-energy tensor and charge current in different frames (both Eckart and Landau frames). This construction allows us to express the fluid in terms of thermodynamic quantities and explicitly extract the transport coefficients. Finally, we formulate the fluid equations of motion in both frames, in terms of these variables and discuss their equivalence with the leading order membrane equations.
%%%%%%%%%%%%%%%%%%%%%%%%%%%%%%%%%%%%%%%%%%%%%%%%%%%%%%%%%%%%%%%
\subsubsection{Analysis in Eckart frame}\label{Analysis in Eckart frame}
%%%%%%%%%%%%%%%%%%%%%%%%%%%%%%%%%%%%%%%%%%%%%%%%%%%%%%%%%%%%%%%
In section~\ref{Eckart Frame}, we established the large $D$ membrane paradigm as a relativistic fluid characterised by the stress-energy tensor $T^{(\text{fluid})\mu \nu}$ and force field $f^\mu$. The heat flux in Eq.~\eqref{HeatFluxGeometric} was introduced in terms of geometric quantities. Now, by replacing the geometric terms with the out-of-equilibrium temperature $T'$, chemical potential $\mu'$, and Eckart velocity field $U^\mu$ using Eqs.~\eqref{ExpT'},~\eqref{Exp:mu'} and~\eqref{Eckartvelocity}, we find the stress-energy tensor  and charge current of membrane fluid up to the first subleading order as
\begin{align}
    &T^{\text{fluid}}_{\mu\nu} = \epsilon(T',\mu') U_\mu U_\nu + q_\mu U_\nu + q_\nu U_\mu-\eta (\nabla^\alpha U^\beta + \nabla^\beta U^\alpha) P_{\alpha\mu} P_{\beta\nu}  + \mathcal{O}(1/D),\label{EMFluidEck.Final} \\
    &J^\mu = n(T',\mu') U^\mu+ \mathcal{O}(1/D),\label{J:muFluidEck.Final}
    \end{align}
where
\begin{align}
 &\epsilon(T',\mu') = \frac{T'(1+8\pi\mu'^2)}{4(1-8\pi\mu'^2)},\notag \\
    &n(T',\mu') = \frac{4\pi\mu' T'}{1-8\pi\mu'^2},\notag\\
    &q_{\mu} = \bigg(C_1{\nabla}^\alpha \mu' +C_2 {\nabla}^2 U^\alpha  + C_3 {\nabla}^\alpha T'+C_4 (U \cdot \nabla U^\alpha) \bigg)P_{\alpha \mu} + \mathcal{O}(1/D).\label{Heatfluxfluid}
    \end{align}
Here, $ C_1, C_2, C_3, C_4,\eta$ are the transport coefficients of the fluid in the Eckart frame, and can be expressed as follows:
    \begin{align}
   &C_1= \frac{(1-Q^2+2Q^4)}{4\sqrt{2\pi} Q}=\frac{(1-8\pi \mu'^2+128\pi^2\mu'^4)}{16\pi\mu'}+\mathcal{O}(1/D),\label{C1} \\
    &C_2=-\frac{(Q^2+1)}{16\pi K}=-\frac{(1-64\pi^2\mu'^4)}{64\pi^2T'}+\mathcal{O}(1/D^2),\label{C2}\\
    &C_3=\frac{Q^2}{4K}=\frac{8\pi\mu'^2(1-8\pi\mu'^2)}{16\pi T'}+\mathcal{O}(1/D^2),\label{C3}\\
    &C_4=\frac{(Q^4+1)}{16\pi}=\frac{(1+64\pi^2\mu'^4)}{16\pi}+\mathcal{O}(1/D),\label{C4}\\
    &\eta=\frac{1}{16\pi}.\label{eta1}
\end{align}
 The fluid equation of motion (Eq.~\eqref{FluidEMEq}) and the charge conservation equation (Eq.~\eqref{FluidChargeEq}) can be written in terms of fluid variables.
The energy conservation equation is
\begin{equation*}
U^\mu \nabla_\mu \epsilon + \epsilon \nabla_\mu U^\mu + \nabla_\mu q^\mu + q_\mu (U^\nu \nabla_\nu U^\mu) -\frac{1}{16\pi} [ (\nabla^\alpha U^\beta + \nabla^\beta U^\alpha) P_\alpha^\mu P_\beta^\nu] \nabla_\mu U_\nu = -U_\mu f^\mu,
\end{equation*}
at leading order in $1/D$ the contribution comes from
\begin{equation}
U^\mu \nabla_\mu \epsilon + \epsilon \nabla_\mu U^\mu + \nabla_\mu q^\mu  -\frac{1}{16\pi} [ (\nabla^\alpha U^\beta + \nabla^\beta U^\alpha) P_\alpha^\mu P_\beta^\nu] \nabla_\mu U_\nu +U_\mu f^\mu = 0.\label{EnergyConEckart}
\end{equation}
The momentum conservation equation is
\begin{align*}
&\epsilon U^\nu \nabla_\nu U^\mu + P^\mu_{\nu} (U^\alpha \nabla_\alpha q^\nu) -\frac{1}{16\pi}  P^\mu_{\nu} \nabla_\alpha[ (\nabla^\beta U^\gamma + \nabla^\gamma U^\beta) P_\beta^\nu P_\gamma^\alpha ]  + q^\alpha \nabla_\alpha U^\mu \\&\quad + q^\mu \nabla_\alpha U^\alpha = P_\nu^\mu f^\nu,
\end{align*}
at leading order in $1/D$ the contribution comes from
\begin{equation}
\epsilon U^\nu \nabla_\nu U^\mu -\frac{1}{16\pi}  P^\mu_{\;\;\nu} \nabla_\alpha[ (\nabla^\beta U^\gamma + \nabla^\gamma U^\beta) P_\beta^\nu P_\gamma^\alpha ] = P_\nu^\mu f^\nu.\label{MomentumConEckart}
\end{equation}
The charge conservation equation is
\begin{equation}
U^\mu \nabla_\mu n + n \nabla_\mu U^\mu = 0.\label{ChargeConEckart}
\end{equation}
 In Appendix~\ref{A2}, we have explicitly shown that the leading order limit of these equations reproduces membrane equations. As the large $D$ expansion in gravity and the derivative expansion in hydrodynamics are independent of each other, our effective membrane fluid generates a second-order derivative term ($\nabla^2U^\alpha$) in the expression of heat flux at the leading order in 1/D analysis. The transport coefficient $C_1$ is related to the energy flow due to the chemical potential gradient, where $C_2$ is the transport coefficient related to the second order velocity change, while $C_3$ is associated with the energy flow solely due to the temperature gradient, the coefficient $C_4$ is related to the energy flow due to the acceleration (which is sourced by the background geometric force), and $\eta$ is the shear viscosity. For membrane fluid, the ratio $\eta/s=1/{4\pi}$. It is important to note that in our basis $C_3$ has a different sign at the leading order from that in a conventional hydrodynamic framework. It is also crucial to emphasise that this hydrodynamic formulation is defined exclusively for a charged membrane ($|Q| > 0$), because the Eckart frame is intrinsically tied to the flow of the conserved charge current. Consequently, the leading order coefficient $C_3 = Q^2/4K$ is nonzero in the parameter space under consideration. As this leading order term cannot vanish within the permissible parameter space, the subleading order corrections remain suppressed, which guarantees that higher-order contributions cannot invert the sign of this transport coefficient at large $D$. So here, it remains strictly positive, leading the heat to flow from the cooler to the hotter region\footnote{A strictly positive transport coefficient associated with the temperature gradient ($C_3 > 0$) dictates that heat flows parallel to the gradient, resulting in a negative effective thermal conductivity.}. A detailed analysis is conducted on this in section~\ref{Thermodynamic Stability}.
 
 In the standard first-order hydrodynamic framework, the relations for heat flux in the Eckart frame typically address only one transport coefficient. This reduction occurs because the three first order (in derivative) variables—the temperature gradient, the chemical potential gradient, and the fluid acceleration—are not independent of each other. As demonstrated in~\cite{Jaiswal:2015mxa, KumarSingh:2025kml}, the momentum conservation equation constrains three of them; the acceleration is driven by the thermodynamic pressure gradient at the leading derivative expansion. By invoking the Gibbs-Duhem relation, this pressure gradient, as well as the acceleration itself, can be expressed as a linear combination of the temperature and chemical potential gradients. This combination leads to one independent transport coefficient - the thermal conductivity. In contrast, the membrane fluid has fundamentally no pressure contribution in the equation of motion and does not obey the Gibbs-Duhem relation at leading order in $1/D$ (as discussed in section~\ref{A4}). The momentum conservation equation (Eq.~\eqref{MomentumConEckart}) demonstrates that the fluid's acceleration ($U\cdot\nabla U^\alpha$) depends on the force term up to first order in the derivative, which completely decouples it from internal pressure gradients. Because the standard algebraic dependence between thermodynamic gradients and acceleration is broken, at first derivative order, the heat flux of the membrane fluid is characterised by three independent transport coefficients ($C_1, C_3, C_4$).
%%%%%%%%%%%%%%%%%%%%%%%%%%%%%%%%%%%%%%%%%%%%%%%%%%%%%%%%%%%%%%%
\subsubsection{Analysis in Landau frame}\label{Analysis in Landau frame}
%%%%%%%%%%%%%%%%%%%%%%%%%%%%%%%%%%%%%%%%%%%%%%%%%%%%%%%%%%%%%%%%
In continuation of section \ref{Landau Frame}, by substitution of geometric quantities ($Q, K $, $u^\mu$) by out-of-equilibrium thermodynamic fluid variables (temperature $T'$, chemical potential $\mu'$, and Landau velocity field $\tilde{U}^\alpha$) in the Landau frame, using Eqs.~\eqref{NumberDiffCurrGeo},~\eqref{LandauVelocity},~\eqref{ExpT'} and~\eqref{Exp:mu'}, we get the form of stress-energy tensor and charge current  as 
\begin{align}
    &T^{\text{fluid}}_{\mu\nu} = \epsilon(T',\mu') \tilde{U}_\mu \tilde{U}_\nu -\eta\left( \nabla^\alpha \tilde{U}^\beta + \nabla^\beta \tilde{U}^\alpha \right) P_{\alpha\mu} P_{\beta\nu} + \mathcal{O}(1/D),\label{EMFluidLandauFinal} \\
    &J^\mu = n(T',\mu') \tilde{U}^\mu + N^\mu + \mathcal{O}(1/D), \label{J:muFluidLandauFinal}
    \end{align}
where
\begin{align}
&\epsilon(T',\mu') = \frac{T'(1+8\pi\mu'^2)}{4(1-8\pi\mu'^2)},\notag \\
    &n(T',\mu') = \frac{4\pi\mu' T'}{1-8\pi\mu'^2},\notag\\
    &N_{\mu} = \bigg( D_1 \nabla^\alpha \mu' +D_2\nabla^2\tilde{U}^\alpha 
    + D_3 \nabla^\alpha T' + D_4 \tilde{U}^c \nabla_c\tilde{U}^\alpha \bigg) P_{\alpha \mu} + \mathcal{O}(1/D).\label{ChargeDiffFinal}
    \end{align}
Here, $D_1,D_2,D_3,D_4,\eta$ are the transport coefficients in the Landau frame, and they take the following forms:
    \begin{align}
    &D_1=-\frac{(1-Q^2+2Q^4)}{(1+Q^2)}=-\frac{(1-8\pi \mu'^2+128\pi^2\mu'^4)}{1+8\pi\mu'^2}+\mathcal{O}(1/D),\label{D1} \\
    &D_2=\frac{Q}{2\sqrt{2\pi}K}=\frac{\mu'(1-8\pi\mu'^2)}{4\pi T'}+\mathcal{O}(1/D^2),\label{D2}\\
    &D_3=-\frac{\sqrt{2\pi}Q^3}{K(1+Q^2)}=-\frac{8\pi(1-8\pi\mu'^2)\mu'^3}{T'(1+8\pi\mu'^2)}+\mathcal{O}(1/D^2),\label{D3}\\
    &D_4=-\frac{Q(1+Q^4)}{2\sqrt{2\pi}(1+Q^2)}=-\frac{(1+64\pi^2\mu'^4)\mu'}{1+8\pi\mu'^2}+\mathcal{O}(1/D),\label{D4}\\
    &\eta=\frac{1}{16\pi}.\label{eta2}
\end{align}
Now, we can describe the conservation equations (which are actually the membrane equations; see Appendix~\ref{A2} for more details) in terms of fluid variables in the Landau frame.  The energy conservation equation is
\begin{equation}
\tilde{U}^\mu \nabla_\mu \epsilon + \epsilon \nabla_\mu \tilde{U}^\mu -\frac{1}{16\pi} [ (\nabla^\alpha \tilde{U}^\beta + \nabla^\beta \tilde{U}^\alpha) P_\alpha^\mu P_\beta^\nu ] \nabla_\mu \tilde{U}_\nu = -\tilde{U}_\mu f^\mu. \label{EnergyConLandau}
\end{equation}
The momentum conservation equation can be described as
\begin{equation}
\epsilon \tilde{U}^\nu \nabla_\nu \tilde{U}^\mu  -\frac{1}{16\pi} P^\mu_{\;\;\nu} \nabla_\alpha [ (\nabla^\beta \tilde{U}^\gamma + \nabla^\gamma \tilde{U}^\beta) P_\beta^\mu P_\gamma^\alpha ] = P^\mu_{\;\;\nu} f^\nu.\label{MomentumConLandau}
\end{equation}
The charge conservation equation takes the following form:
\begin{equation}
\tilde{U}^\mu \nabla_\mu n + n \nabla_\mu\tilde{U}^\mu + \nabla_\mu N^\mu = 0.\label{ChargeConLandau}
\end{equation}
Similar to the analysis in the Eckart frame, the membrane fluid generates a second-order derivative term ($ \nabla^2 \tilde{U}^\alpha$) in the leading order $1/D$ analysis in the Landau frame. Because the acceleration of this fluid is sourced from the geometric force (Eq.~\eqref{MomentumConLandau}), and it does not obey the thermodynamic Gibbs-Duhem relation at leading order in $1/D$, the charge diffusion current is characterised here by three independent first order transport coefficients: $D_1$ (associated with the chemical potential gradient), $D_3$ (associated with the temperature gradient), and $D_4$ (associated with acceleration). Additionally, $D_2$ is a purely second-order transport coefficient related to the second-order derivative of the Landau velocity field. As in the Eckart frame, the shear viscosity $\eta$ satisfies the relation $\eta/s = 1/4\pi$.
It is also important to note that, similar to $C_3$ in the Eckart frame, the coefficient $D_3$ has a different sign from that in conventional hydrodynamics\footnote{In usual hydrodynamics, the coefficient of the temperature gradient in the expression of charge diffusion current is positive for positive chemical potential and negative for negative chemical potential.}. Specifically, for positive chemical potential ($\mu' \propto Q > 0$), $D_3$ is strictly negative, as is visible from Eq.~\eqref{D3}. Since the velocity traces the energy flow in the Landau frame, there is no energy flux, and dissipation manifests as a number- or charge-diffusion current. In Eq.~\eqref{ChargeDiffFinal}, the negative coefficient multiplying the temperature gradient in the charge diffusion current indicates that charge diffusion occurs from a high temperature region to a low temperature region. Similarly, if we consider a negative chemical potential ($Q<0$), the coefficient $D_3$ is strictly positive, which implies that charge diffusion occurs from a low temperature to a high temperature region. As discussed earlier for $C_3$, the leading-order coefficient $D_3$ is non-zero for a charged membrane ($|Q|>0$), and no subleading order correction can reverse its sign. Moreover, while $D_3$ strictly governs charge transport rather than heat flux in this frame, the effective thermal conductivity is instead characterised by $C_3$, as discussed in Section~\ref{Thermodynamic Stability}.
%%%%%%%%%%%%%%%%%%%%%%%%%%%%%%%%%%%%%%%%%%%%%%%%%%%%%%%%%%%%%%%
\section{Thermodynamic stability}\label{Thermodynamic Stability}
%%%%%%%%%%%%%%%%%%%%%%%%%%%%%%%%%%%%%%%%%%%%%%%%%%%%%%%%%%%%
The equations of motion of the fluid are dynamically equivalent to the membrane equations at the leading order in $1/D$. In Ref.~\cite{Bhattacharyya:2015fdk}, the light quasinormal mode analysis—evaluated by perturbing the membrane velocity $u^\mu$, shape $K$, and charge $Q$ in the linearised membrane equations around a membrane configuration corresponding to a RN black hole—rigorously demonstrates the membrane's stability. Because the Reissner-Nordström solution maps precisely to the global thermodynamic equilibrium of the dual fluid (as discussed in section~\ref{Global Thermodynamic Equilibrium}), and the local fluid variables (temperature, chemical potential, and fluid velocity) are exact algebraic combinations of the geometric membrane data (shape, charge density, and membrane velocity) defined in section~\ref{Local Thermodynamic Equilibrium}, the fluid conservation equations are mathematically isomorphic to the membrane equations. Consequently, evaluating the decay of all local hydrodynamic perturbations by linearising the fluid equations is algebraically indistinguishable from linearising the membrane equations. This exact mathematical equivalence guarantees that if the thermodynamic and kinematic quantities are perturbed from an equilibrium state, the perturbations will damp out over time exactly as dictated by the quasinormal modes, restoring global thermodynamic equilibrium and rendering an independent fluid stability calculation redundant.

Consider a fluid element of volume $\delta v$. Geometrically, this volume corresponds to an area element $\delta A$ on the membrane hypersurface. As discussed in section~\ref{Large D membrane paradigm}, membrane paradigm is constructed by patching local Reissner-Nordström solutions together implying the local area scales as $\delta A \propto (r_0(x))^{(D-2)}$, which also implies an identical scaling for the fluid volume element, $\delta v \propto r_0(x)^{D-2}$. Eqs.~\eqref{Redefinition2}, and~\eqref{Fluid:epsilon} demonstrate that the leading order energy density scales  as  $\epsilon \propto 1/r_0(x)$. Then the total energy of a fluid occupying volume $\delta v$ is
    $$E=\epsilon\delta v \propto (r_0(x))^{(D-3)}.$$
Eqs.~\eqref{ExpT'} and~\eqref{Temp} establish that the leading order local temperature scales as follows,
$$ T'\propto 1/r_0(x).$$
Evaluating the heat capacity for constant chemical potential (constant Q) as $$C = \left(\frac{\partial E}{\partial T'}\right)_{\mu'} = \frac{\partial E / \partial r_0}{\partial T' / \partial r_0} < 0.$$

The fluid element exhibits negative heat capacity, which means that an influx of energy decreases its temperature, while an outflux of energy increases it. Eq.~\eqref{Heatfluxfluid} and Eq.~\eqref{C3} demonstrate that the transport coefficient for the temperature gradient in the expression of heat flux is strictly positive, leading to negative effective thermal conductivity, implying heat flows from the lower-temperature to the higher-temperature region. In this specific representation, driven solely by temperature gradient, heat moves from a low-temperature region to a high-temperature region, cooling the hotter region and heating the cooler region. This phenomenon lowers the temperature gradient and drives the system towards global thermal equilibrium, as suggested by the membrane's quasi-normal mode analysis. Note that this negative heat capacity is unaffected by subleading corrections. Since the geometric scaling of the fluid's volume dictates that the energy $E \propto r_0^{D-3}$, while the temperature scales as $T' \propto 1/r_0$, the negative sign of the derivative is driven by a leading $\mathcal{O}(D)$ scaling. Therefore, $\mathcal{O}(1)$ corrections are suppressed and cannot reverse the sign.
%%%%%%%%%%%%%%%%%%%%%%%%%%%%%%%%%%%%%%%%%%%%%%%%%%%%%%%%%%%%%%%%%%
\section{Summary and outlook}\label{Summary and outlook}
%%%%%%%%%%%%%%%%%%%%%%%%%%%%%%%%%%%%%%%%%%%%%%%%%%%%%%%%%%%%%%%%%%
In summary, we have established a precise hydrodynamic duality for the charged large $D$ membrane in an asymptotically flat background. The geometric evolution of the membrane at leading order in 1/D reduces to the dynamics of a relativistic charged fluid driven by an effective background force arising from the extrinsic curvature ($K_{\mu\nu}$) of the membrane hypersurface (Eq.~\eqref{Forceform}). After explicitly calculating the thermodynamic variables, we have formulated the fluid system in distinct frames and have evaluated the non-equilibrium dynamics. Eqs.~\eqref{C1}-\eqref{eta1} demonstrate the transport coefficients of the membrane fluid in the Eckart frame, whereas Eqs.~\eqref{D1}-\eqref{eta2} represent them in the Landau frame. Further, we demonstrated that the thermodynamic stability of the fluid is governed by an unconventional mechanism: a negative heat capacity coupled with a negative effective thermal conductivity. This mechanism quenches local thermal gradients. The system dynamically dampens perturbations and ensures the system's relaxation back to global thermodynamic equilibrium, consistent with gravitational quasinormal mode analysis.

In the current framework, the stress-energy tensor and charge current incorporate only specific parts of the subleading order (in $1/D$) corrections. Consequently, the resulting effective theory captures only a single second-order (in derivative) hydrodynamic term. A comprehensive derivation of the second-order fluid dynamics, including the explicit reading off of the second-order transport coefficients, requires extending this formalism for the subleading order (in $1/D$) membrane equations.  While constructing the leading order (in $1/D$) fluid equation, we get an unconventional hydrodynamic structure in the stress-energy tensor. Because of this, the standard phenomenological construction of a local hydrodynamic entropy current is not possible here; it must be derived directly within the large $D$ membrane paradigm. We also note that computing the expression of membrane entropy current $J_s^\mu$ constitutes a distinct analytical challenge. While Hawking’s area theorem guarantees global positive-definite entropy production, demonstrating local entropy production ($\nabla_\mu J_s^\mu \ge 0$) within the fluid requires a systematic derivation of it. We intend to explore these aspects in future work. Furthermore, future work will aim to construct a fluid dual to an asymptotically flat black hole using the large $D$ membrane paradigm such that it resides on the asymptotic null boundary of the black hole, providing a picture analogous to AdS/Hydrodynamics, where, since the boundary is a null manifold, such a fluid, if it can be constructed, has to be Carrollian.
%%%%%%%%%%%%%%%%%%%%%%%%%%%%%%%%%%%%%%%%%%%%%%%%%%%%%%%%%%%%%%%%%%%%%%%%%%%%%%%%%%%%%%%%
\acknowledgments
%%%%%%%%%%%%%%%%%%%%%%%%%%%%%%%%%%%%%%%%%%%%%%%%%%%%%%%%%%%%%%%
We would like to thank the people of India for their continuous support to the scientific research in India. S.H. acknowledges the financial support from the Institute Fellowship provided by the Indian Institute of Technology (ISM) Dhanbad. M.K. would like to acknowledge the Department of Science and Technology (DST), Govt. of India, for the INSPIRE-Faculty award (DST/INSPIRE/04/2024/001794), and the Faculty Research Scheme (FRS project number: MISC 0240) at IIT (ISM) Dhanbad. The authors additionally acknowledge the Department of Physics at the Indian Institute of Technology (ISM) Dhanbad for providing the necessary academic infrastructure.
\appendix
%%%%%%%%%%%%%%%%%%%%%%%%%%%%%%%%%%%%%%%%%%%%%%%%
\section{Detailed calculations }
%%%%%%%%%%%%%%%%%%%%%%%%%%%%%%%%%%%%%%%%%%%%%
\subsection{Calculation for Table 1}\label{A1}
%%%%%%%%%%%%%%%%%%%%%%%%%%%%%%%%%%%%%%%%%%%%%%
Using Eq.~\eqref{EMAppendix} the exact calculation to derive the quantities listed in Table~\ref{table1} is performed by the following steps:
\begin{align}
    T_{\mu\nu} &= s_1 u_\mu u_\nu - \frac{v_\mu}{8\pi} u_\nu - \frac{v_\nu}{8\pi} u_\mu + w_{\mu\nu}+\mathcal{O}(1/D), \notag \\
    &= s_1 u_\mu u_\nu - \frac{g_{\mu\alpha} v^\alpha}{8\pi} u_\nu - \frac{g_{\nu\beta} v^\beta}{8\pi} u_\mu + w^{\alpha\beta} g_{\mu\alpha} g_{\nu\beta} +\mathcal{O}(1/D),\notag \\
    &= s_1 u_\mu u_\nu - \frac{(P_{\mu\alpha} - u_\mu u_\alpha) v^\alpha}{8\pi} u_\nu - \frac{(P_{\nu\beta} - u_\nu u_\beta) v^\beta}{8\pi} u_\mu\notag \\&\quad+ w^{\alpha\beta} (P_{\mu\alpha} - u_\mu u_\alpha) (P_{\nu\beta} - u_\nu u_\beta)+\mathcal{O}(1/D), \notag \\
    &= \biggl( s_1 + \frac{v^\alpha u_\alpha}{4\pi} + w^{\alpha\beta} u_\alpha u_\beta \biggr) u_\mu u_\nu + \biggl( -\frac{1}{8\pi} v^\alpha P_{\mu\alpha} - w^{\alpha\beta} P_{\mu\alpha} u_\beta \biggr) u_\nu \notag \\
    &\quad + \biggl( -\frac{1}{8\pi} v^\beta P_{\nu\beta} - w^{\alpha\beta} P_{\nu\beta} u_\alpha \biggr) u_\mu + w^{\alpha\beta} P_{\mu\alpha} P_{\nu\beta} +\mathcal{O}(1/D),\notag \\
    &= \epsilon u_\mu u_\nu + l_\mu u_\nu + l_\nu u_\mu + r_{\mu\nu}+\mathcal{O}(1/D),\label{T:mu:nu} 
\end{align}
where
\begin{equation}\notag
    \begin{cases}
        \epsilon = s_1 + \frac{v^\alpha u_\alpha}{4\pi} + w^{\alpha\beta} u_\alpha u_\beta \,, \\[1.5ex]
        l_\mu = -\frac{v^\alpha P_{\mu\alpha}}{8\pi} - w^{\alpha\beta} P_{\mu\alpha} u_\beta \,, \\[1.5ex]
        r_{\mu\nu} = w^{\alpha\beta} P_{\mu\alpha} P_{\nu\beta} .
    \end{cases}
\end{equation}
Substituting the forms of $s_1$, $v^\alpha$, and $w^{\alpha\beta}$ in terms of membrane data using Eq.~\eqref{expv} and Eq.~\eqref{Exps1w}, in the above expression, we get:
\begin{align}
    \epsilon &= \frac{K}{16\pi} (1 + Q^2) + \frac{1}{4\pi} \bigl[ Q u \cdot \nabla Q + Q^2 u \cdot K \cdot u  \notag \\
    &\quad + \biggl( \frac{2Q^4 - Q^2 - 1}{2} \biggr) \frac{u \cdot \nabla K}{K} + \biggl( \frac{1+Q^2}{K} \biggr) u_\alpha \nabla^2 u^\alpha\bigr] + \biggl( \frac{1-Q^2}{16\pi} \biggr) u \cdot K \cdot u, \notag \\
    &= \frac{K}{16\pi} (1 + Q^2) + \frac{1}{8\pi} \biggl[ \biggl( \frac{1+3Q^2}{2} \biggr) u \cdot K \cdot u + 2 Q u \cdot \nabla Q \notag \\
    &\quad + \biggl( \frac{2Q^4 - Q^2 - 1}{K} \biggr) u \cdot \nabla K \biggr] .\label{Expe}
\end{align}
Here, we have used the fact that $u_\alpha \nabla^2 u^\alpha \sim \mathcal{O}(1)$ (for detail calculation see Ref.~\cite{Bhattacharyya:2016nhn}).
\begin{align}
    l_\mu &= -\frac{1}{8\pi} \biggl[ Q \nabla^\alpha Q + Q^2 u_\beta K^{\alpha\beta} + \biggl( \frac{2Q^4 - Q^2 - 1}{2} \biggr) \frac{\nabla^\alpha K}{K} \notag \\
    &\quad - \biggl( \frac{Q^2 + 2Q^4}{2} \biggr) u \cdot \nabla u^\alpha + \biggl( \frac{1+Q^2}{K} \biggr) \nabla^2 u^\alpha \biggr] P_{\alpha\mu} \notag \\
    &\quad - \frac{1}{8\pi} \biggl[ \biggl( \frac{1-Q^2}{2} \biggr) u_\beta K^{\alpha\beta} - \frac{1}{2} u \cdot \nabla u^\alpha \biggr] P_{\alpha\mu}, \notag \\
    &= -\frac{1}{8\pi} \biggl[ Q \nabla^\alpha Q + \biggl( \frac{1+Q^2}{2} \biggr) u_\beta K^{\alpha\beta} + \biggl( \frac{2Q^4 - Q^2 - 1}{2} \biggr) \frac{\nabla^\alpha K}{K} \notag \\
    &\quad + \biggl( \frac{1+Q^2}{K} \biggr) \nabla^2 u^\alpha - \biggl( \frac{1+Q^2 + 2Q^4}{2} \biggr) u \cdot \nabla u^\alpha \biggr] P_{\alpha\mu} ,\label{Exp:l:mu}
\end{align}
and finally, we obtain
\begin{equation}
    r_{\mu\nu} = \frac{1}{16\pi} \bigl[ (1-Q^2) K^{\alpha\beta} - (\nabla^\alpha u^\beta + \nabla^\beta u^\alpha) \bigr] P_{\alpha\mu} P_{\beta\nu}. \label{Expr:mu:nu}
\end{equation}

%%%%%%%%%%%%%%%%%%%%%%%%%%%%%%%%%%%%%%%%%%%%%%%%%%%%%
\subsection{Consistency of hydrodynamic frame transformations}\label{A0}
%%%%%%%%%%%%%%%%%%%%%%%%%%%%%%%%%%%%%%%%%%%%%%%%%%%%%%
From Eqs.~\eqref{LandauFluidEM}-\eqref{ExpN_a}, we obtain
\begin{align}
   & T_{\mu\nu}^{fluid} = \epsilon \tilde{U}_\mu \tilde{U}_\nu - \frac{1}{16\pi} (\nabla^\alpha \tilde{U}^\beta + \nabla^\beta \tilde{U}^\alpha) P_{\alpha\mu} P_{\beta\nu} + \mathcal{O}(1/D),\\
   &J_\mu = n \tilde{U}_\mu - n \frac{q_\mu}{\epsilon}.
\end{align}
To verify the consistency of the frame transformation using Eq.~\eqref{LandauVelocity}, the substitution of $\tilde{U}_\mu$ in the above equations yields
\begin{align*}
    T_{\mu\nu}^{fluid} &= \epsilon \left(U_\mu + \frac{q_\mu}{\epsilon}\right) \left(U_\nu + \frac{q_\nu}{\epsilon}\right) \nonumber \\
    &\quad - \frac{1}{16\pi} \left[ \nabla^\alpha \left(U^\beta + \frac{q^\beta}{\epsilon}\right) + \nabla^\beta \left(U^\alpha + \frac{q^\alpha}{\epsilon}\right) \right] P_{\mu\alpha} P_{\nu\beta} + \mathcal{O}(1/D) \\
    &= \epsilon U_\mu U_\nu + U_\mu q_\nu + U_\nu q_\mu - \frac{1}{16\pi} (\nabla^\alpha U^\beta + \nabla^\beta U^\alpha) P_{\alpha\mu} P_{\beta\nu} + \mathcal{O}(1/D),\\
      J_\mu &= n \left(U_\mu + \frac{q_\mu}{\epsilon}\right) - n \frac{q_\mu}{\epsilon} + \mathcal{O}(1/D) \\
    &= n U_\mu + \mathcal{O}(1/D).
\end{align*}
With the above substitution we successfully reproduce Eqs.~\eqref{Eck.Fluid.EM} and~\eqref{EckartChargeCurrent} up to $\mathcal{O}(1)$.
%%%%%%%%%%%%%%%%%%%%%%%%%%%%%%%%%%%%%%%%%%%%%%%%%%%%%
\subsection{Out-of-equilibrium redefinitions}\label{A2}
%%%%%%%%%%%%%%%%%%%%%%%%%%%%%%%%%%%%%%%%%%%%%%%%%%%%%%
From section~\ref{Local Thermodynamic Equilibrium}, we have the following relations:
\begin{align}
    &\epsilon(T',\mu')=\epsilon(T,\mu) + \frac{1}{8\pi} \bigg[  \frac{(3Q^2+1)}{2} u^\alpha u^\beta K_{\alpha\beta} + 2Q u \cdot \nabla Q + \frac{(2Q^4 - Q^2 - 1)}{K} u \cdot \nabla K   \bigg],\label{Redefinition1A} \\
    &n(T',\mu')=n(T,\mu).\label{Redefinition2A}
\end{align}
As $\delta T = \mathcal{O}(1)$, $\delta\mu = \mathcal{O}(1/D)$, and $T = \mathcal{O}(D)$, $\mu=\mathcal{O}(1)$. We expand the expressions and retain the terms upto $\mathcal{O}(1)$  and neglect higher order terms in $1/D$, such as $\delta\mu\delta T \sim \mathcal{O}(1/D)$ and $(\delta\mu)^2 \sim \mathcal{O}(1/D^2)$, as per Eq.~\eqref{Redefinition2A}: 
\begin{align}
&n(T',\mu')=n(T,\mu),\notag\\
\Rightarrow& \frac{4\pi\mu T}{1 - 8\pi\mu^2} = \frac{4\pi(\mu + \delta\mu)(T + \delta T)}{1 - 8\pi(\mu + \delta\mu)^2},\notag \\
\Rightarrow &\mu T (1 - 8\pi\mu^2) - 16\pi\mu^2 T \delta\mu = (1 - 8\pi\mu^2) [\mu T + T \delta\mu + \mu \delta T],\notag \\
\Rightarrow & -16\pi\mu^2 T \delta\mu - (1 - 8\pi\mu^2) T \delta\mu = (1 - 8\pi\mu^2) \mu \delta T,\notag \\
\Rightarrow &\delta\mu = -\frac{\mu}{T} \frac{1 - 8\pi\mu^2}{1 + 8\pi\mu^2} \delta T.\label{del:mu}
\end{align}
We define the term $\mathcal{L}$ as
\begin{equation*}
\mathcal{L} \equiv \left[ \frac{(3Q^2+1)}{2}  u^\alpha u^\beta K_{\alpha\beta} + 2Q u \cdot \nabla Q + \frac{2Q^4 - Q^2 - 1}{K} u \cdot \nabla K \right].
\end{equation*}
Applying the same $\mathcal{O}(1/D)$ truncation, we expand the terms of Eq.~\eqref{Redefinition1A}, which yields the following:
\begin{align}
&\frac{T + \delta T}{4} \frac{1 + 8\pi(\mu + \delta\mu)^2}{1 - 8\pi(\mu + \delta\mu)^2} = \frac{T}{4} \frac{1 + 8\pi\mu^2}{1 - 8\pi\mu^2} + \frac{\mathcal{L}}{8\pi},\notag \\[10pt]
\Rightarrow& (1 - 8\pi\mu^2) T (16\pi\mu \delta\mu) + \delta T [1 + 8\pi(\mu^2 + 2\mu\delta\mu)](1 - 8\pi\mu^2)\notag \\
&\qquad = (1 + 8\pi\mu^2) T (-16\pi\mu\delta\mu) + \frac{1}{2\pi}(1 - 8\pi\mu^2)^2 \mathcal{L},\notag \\[10pt]
\Rightarrow& 32\pi\mu T \delta\mu + \delta T (1 + 8\pi\mu^2)(1 - 8\pi\mu^2) = \frac{1}{2\pi}(1 - 8\pi\mu^2)^2 \mathcal{L},\notag \\[10pt]
\Rightarrow& -32\pi\mu^2 \frac{1 - 8\pi\mu^2}{1 + 8\pi\mu^2} \delta T + \delta T (1 + 8\pi\mu^2)(1 - 8\pi\mu^2) = \frac{1}{2\pi}(1 - 8\pi\mu^2)^2 \mathcal{L}, \notag\\[10pt]
\Rightarrow& \delta T (1 + 8\pi\mu^2)^2 - 32\pi\mu^2 \delta T = \frac{1}{2\pi}(1 - 8\pi\mu^2)(1 + 8\pi\mu^2) \mathcal{L},\notag \\[10pt]
\Rightarrow& \delta T (1 - 8\pi\mu^2)^2 = \frac{1}{2\pi}(1 - 8\pi\mu^2)(1 + 8\pi\mu^2) \mathcal{L},\notag\\
\Rightarrow& \delta T = \frac{1 + 8\pi\mu^2}{2\pi(1 - 8\pi\mu^2)} \mathcal{L}.\label{delta:T}
\end{align}
Substituting Eq.~\eqref{delta:T} to Eq.~\eqref{del:mu} we get
\begin{align}
 \delta\mu = -\frac{\mu \mathcal{L}}{2\pi T}.
\end{align}
By substituting the expression of $\mathcal{L}$, $\mu$ and $T$ in terms of geometric quantities, our final expressions of temperature and chemical potential correction take the following forms:
\begin{align*}
    &\delta T= \frac{1+Q^2}{2\pi(1-Q^2)} \bigg[ \frac{(3Q^2+1)}{2}  u^\alpha u^\beta K_{\alpha\beta} + 2Q u \cdot \nabla Q + \frac{(2Q^4 - Q^2 - 1)}{K} u \cdot \nabla K   \bigg],\\ 
    &\delta\mu=\frac{-Q}{\sqrt{2\pi}K(1-Q^2)} \bigg[  \frac{(3Q^2+1)}{2}  u^\alpha u^\beta K_{\alpha\beta} + 2Q u \cdot \nabla Q + \frac{(2Q^4 - Q^2 - 1)}{K} u \cdot \nabla K   \bigg],
\end{align*}
respectively.
%%%%%%%%%%%%%%%%%%%%%%%%%%%%%%%%%%%%%%%%%%%%%%%
\subsection{Fluid equation to membrane equation}\label{A3}
%%%%%%%%%%%%%%%%%%%%%%%%%%%%%%%%%%%%%%%%%%%%%%%%%
In continuation of section~\ref{Hydrodynamic equations and constitutive relations}, here we explicitly demonstrate the calculation for the leading order (in $1/D$) equivalence between the fluid equations and the membrane equations. Using Eqs.~\eqref{MomentumConEckart},~\eqref{EnergyConEckart}, and~\eqref{ChargeConEckart} we demonstrate the following steps:
%%%%%%%%%%%%%%%%%%%%%%%%%%%%%%%%%%%%%%%%%%%%%%%%%
\subsection*{Momentum conservation relation}
%%%%%%%%%%%%%%%%%%%%%%%%%%%%%%%%%%%%%%%%%%%%%%%%
\begin{align*}
&\epsilon U^\nu \nabla_\nu U^\mu - \frac{1}{16\pi} P^\mu_\nu \nabla_\alpha \left[ (\nabla^\beta U^\gamma + \nabla^\gamma U^\beta) P^\nu_\beta P^\alpha_\gamma \right] = P^\mu_\nu f^\nu, \\
\Rightarrow{} & \epsilon U^\nu \nabla_\nu U^\mu - \frac{1}{16\pi} P^\mu_\nu \nabla_\alpha \left[ (\nabla^\beta U^\alpha + \nabla^\alpha U^\beta) P^\nu_\beta \right] + \frac{1-Q^2}{16\pi} P^\mu_\nu \nabla_c (K^{ab} P^c_a P^\nu_b) +\mathcal{O}(1) = 0, \\
\Rightarrow{} & \epsilon U^\nu \nabla_\nu U^\mu - \frac{1}{16\pi} P^\mu_\nu \nabla_\alpha \left[ \nabla^\nu U^\alpha + \nabla^\alpha U^\nu \right] \\
& + \frac{1-Q^2}{16\pi} P^\mu_\nu \nabla_c (K^{c\nu} + K^{cb} U_b U^\nu + K^{\nu b} U_c U_a) +\mathcal{O}(1) = 0, \\
\Rightarrow{} & \frac{K(1+Q^2)}{16\pi} u^\nu \nabla_\nu u^\mu - \frac{1}{16\pi} P^\mu_\nu (\nabla^2 u^\nu + \nabla_\alpha \nabla^\nu u^\alpha) + \frac{1-Q^2}{16\pi} P^\mu_\nu \nabla_c K^{c\nu} +\mathcal{O}(1) = 0,
\end{align*}
\begin{equation}\label{MomCon}
    \Rightarrow{} 
 (1 + Q^2)(u . \nabla) u^\mu + (1 - Q^2)P^{\mu\nu} \left( \frac{\nabla _\nu K}{K} \right) - P^{\mu\nu} \left( \frac{\nabla^2 u_\nu}{K} + K_{\nu\alpha}u^\alpha \right)=\mathcal{O}(1/D).
\end{equation}
%%%%%%%%%%%%%%%%%%%%%%%%%%%%%%%%%%%%%%%%%%%%
\subsection*{Energy conservation relation}
%%%%%%%%%%%%%%%%%%%%%%%%%%%%%%%%%%%%%%%%%%%%
\begin{align*}
& U^\mu \nabla_\mu \epsilon + \epsilon \nabla_\mu U^\mu + \nabla_\mu q^\mu - \frac{1}{16\pi} \left[ (\nabla^\alpha U^\beta + \nabla^\beta U^\alpha) P^\mu_\alpha P^\nu_\beta \right] \nabla_\mu U_\nu + U^\mu f_\mu = 0, \\[10pt]
\Rightarrow{} & u^\mu \nabla_\mu \epsilon + \epsilon \left( \nabla_\mu u^\mu + \frac{\nabla_\mu M^\mu}{L} \right) + \nabla_\mu l^\mu - \frac{\epsilon}{L} \nabla_\mu M^\mu - \frac{1-Q^2}{16\pi} \left[ u^\mu \nabla^\nu (K^{ab} P_{a\mu} P_{b\nu}) \right]\\
& + \frac{1}{16\pi} u_\nu \nabla_\mu \left[ (\nabla^\nu u^\mu + \nabla^\mu u^\nu) + u^\nu (u \cdot \nabla u^\mu) + u^\mu (u \cdot \nabla u^\nu) \right]  + \mathcal{O}(1) = 0, \\[10pt]
\Rightarrow{} & u^\mu \nabla_\mu \epsilon + \epsilon\nabla_\mu u^\mu - \nabla_\mu \left( \frac{v_\alpha P^{\mu\alpha}}{8\pi} +w_{\alpha\beta}u^\alpha P^{\mu\beta}\right)- \frac{1-Q^2}{16\pi} \left[ u^\mu \nabla^\nu (K^{ab} P_{a\mu} P_{b\nu}) \right]\\
& + \frac{1}{16\pi} u_\nu \nabla_\mu \left[ (\nabla^\nu u^\mu + \nabla^\mu u^\nu) + u^\nu (u \cdot \nabla u^\mu) + u^\mu (u \cdot \nabla u^\nu) \right] + \mathcal{O}(1) = 0, \\[10pt]
\Rightarrow{} & u^\mu \nabla_\mu \epsilon + \epsilon \nabla_\mu u^\mu - \nabla_\mu \left( \frac{v_\alpha P^{\mu\alpha}}{8\pi} \right) - \frac{1-Q^2}{16\pi} \nabla_\mu (P^{\alpha\mu} u^\beta K_{\alpha\beta}) + \frac{1}{16\pi} \nabla_\mu (u \cdot \nabla u^\mu) \\
&- \frac{1-Q^2}{16\pi} \left[ u^\mu \nabla^\nu (K_{\mu\nu} + u_\mu K^b_\nu u_b) \right] + \frac{1}{16\pi} u_\nu \nabla_\mu (\nabla^\mu u^\nu + \nabla^\nu u^\mu) - \frac{1}{16\pi} \nabla_\mu (u \cdot \nabla u^\mu) \\& +\mathcal{O}(1) = 0, \\[10pt]
\Rightarrow{} & \frac{K u^\mu}{16\pi} \nabla_\mu Q^2 + \frac{1+Q^2}{16\pi} u^\mu \nabla_\mu K + \frac{K(1+Q^2)}{16\pi} \nabla_\mu u^\mu - \frac{1}{8\pi} \nabla_\mu v^\mu - \frac{1-Q^2}{16\pi} \nabla_\mu (u_\beta K^{\beta\mu}) \\
& - \frac{1-Q^2}{16\pi} u_\mu \nabla_\nu K^{\nu\mu}+ \frac{1-Q^2}{16\pi} \nabla^\nu (K^b_\nu u_b) + \frac{1}{16\pi} u_\nu \nabla_\mu (\nabla^\mu u^\nu + \nabla^\nu u^\mu)  +\mathcal{O}(1) = 0, \\[10pt]
\Rightarrow{} & \frac{K}{2} u \cdot \nabla Q^2 + \frac{1+Q^2}{2} u \cdot \nabla K + \frac{K(1+Q^2)}{2} \nabla_\mu u^\mu - \nabla_\mu v^\mu \\
& + \frac{1}{2} u_\nu \nabla_\mu (\nabla^\mu u^\nu + \nabla^\nu u^\mu) - \frac{1-Q^2}{2} u_\mu \nabla_\nu K^{\mu\nu} +\mathcal{O}(1) = 0,
\end{align*}
\begin{equation}\label{EnCon}
\Rightarrow{}\nabla \cdot u=\mathcal{O}(1/D).
\end{equation}
The derivation of the last line from the second last one is explicitly done in \cite{Bhattacharyya:2016nhn}.
Throughout this calculation we have used the following identities:
\begin{align*}
 &\nabla_\nu(u\cdot\nabla u^\nu)=K(u\cdot K\cdot u),\\
 &\nabla_\mu K^\mu_\nu=\nabla_\mu K,\\
 &\nabla_\alpha \nabla_\mu u^\alpha=K(u^\alpha K_{\nu \alpha}),\\
 &\nabla\cdot u=\mathcal{O}(1).
\end{align*}
The systematic proof of these identities can be found in~\cite{Bhattacharyya:2016nhn}.
%%%%%%%%%%%%%%%%%%%%%%%%%%%%%%%%%%%%%%%%%%%%%%
\subsection*{Charge conservation relation}
%%%%%%%%%%%%%%%%%%%%%%%%%%%%%%%%%%%%%%%%%%%%%%%%
\begin{align*}
& n \nabla_\mu U^\mu + U^\mu \nabla_\mu n = 0, \\[10pt]
\Rightarrow{} & n \nabla_\mu u^\mu + n \frac{\nabla_\mu M^\mu}{L} + u^\mu \nabla_\mu n +\mathcal{O}(1) = 0, \\[10pt]
\Rightarrow{} & \nabla_\mu M^\mu + u \cdot \nabla \left( \frac{K Q}{2\sqrt{2\pi}} \right) +\mathcal{O}(1) = 0, \\[10pt]
\Rightarrow{} & \nabla_\mu \left( - \nabla_\alpha Q P^{\alpha\mu} - Q u \cdot \nabla u^\mu \right) + u \cdot \nabla (K Q) + \mathcal{O}(1) = 0,
\end{align*}
\begin{equation}
    \Rightarrow{}  \frac{\nabla^2 Q}{K} +  Q \, u \cdot K \cdot u -\frac{ u \cdot \nabla (K Q)}{K} = \mathcal{O}(1/D).
\end{equation}
In the above calculation, we used Eq.~\eqref{EnCon} to obtain the leading order contribution. 
To prove the same in Landau frame, we substitute the expression of $\tilde{U}_\mu$ and $N_\mu$ in terms of $U_\mu$ and $q_\mu$ using Eq.~\eqref{LandauVelocity} and Eq.~\eqref{ExpN_a} in Eqs.~\eqref{EnergyConLandau},~\eqref{MomentumConLandau}, and~\eqref{ChargeConLandau}. This substitution reproduces Eqs.~\eqref{EnergyConEckart},~\eqref{MomentumConEckart}, and~\eqref{ChargeConEckart} at  the leading order in $1/D$. The proof then follows straightforwardly from the above calculation.
%%%%%%%%%%%%%%%%%%%%%%%%%%%%%%%%%%%%%%%%%%%%%%%%%%%%%

\bibliographystyle{JHEP}
\bibliography{biblio}

\end{document}